\documentclass[
 reprint,
 amsmath,amssymb,
 aps,raggedbottom
]{revtex4-2}

\usepackage{bm}
\usepackage{dcolumn}
\usepackage{graphicx}
\usepackage{mathtools}
\usepackage[T1]{fontenc}

\begin{document}

\preprint{APS/123-QED}

\title{Multifractal emergent processes: Multiplicative interactions override nonlinear component properties}

\author{Madhur Mangalam}
\email{Correspondence should be sent to: \href{mailto:mmangalam@unomaha.edu}{mmangalam@unomaha.edu}.}
\affiliation{Division of Biomechanics and Research Development, Department of Biomechanics, and Center for Research in Human Movement Variability, University of Nebraska at Omaha, NE 68182, USA}

\author{Damian G. Kelty-Stephen}
\email{Correspondence should be sent to: \href{mailto:keltystd@newpaltz.edu}{keltystd@newpaltz.edu}.}
\affiliation{Department of Psychology, State University of New York at New Paltz, New Paltz, New York 12561, USA}

\begin{abstract}
    Biological and psychological processes have been conceptualized as emerging from intricate multiplicative interactions among component processes across various scales. Among the statistical models employed to approximate these intricate nonlinear interactions, one prominent framework is that of cascades. Despite decades of empirical work using multifractal formalisms, a fundamental question has persisted: Do the observed nonlinear interactions across scales owe their origin to multiplicative interactions, or do they inherently reside within the constituent processes? This study presents the results of rigorous numerical simulations that demonstrate the supremacy of multiplicative interactions over the intrinsic nonlinear properties of component processes. To elucidate this point, we conducted simulations of cascade time series featuring component processes operating at distinct timescales, each characterized by one of four properties: multifractal nonlinearity, multifractal linearity (obtained via the Iterative Amplitude Adjusted Wavelet Transform of multifractal nonlinearity), phase-randomized linearity (obtained via the Iterative Amplitude Adjustment Fourier Transform), and phase- and amplitude-randomized (obtained via shuffling). Our findings unequivocally establish that the multiplicative interactions among components, rather than the inherent properties of the component processes themselves, decisively dictate the multifractal emergent properties. Remarkably, even component processes exhibiting purely linear traits can generate nonlinear interactions across scales when these interactions assume a multiplicative nature. In stark contrast, additivity among component processes inevitably leads to a linear outcome. These outcomes provide a robust theoretical underpinning for current interpretations of multifractal nonlinearity, firmly anchoring its roots in the domain of multiplicative interactions across scales within biological and psychological processes.
\end{abstract}


\maketitle

\section{Introduction}

\subsection{Multifractal geometry arising from multiplicative dynamics in biology and psychology}

Conventional approaches in biology and psychology have often sought to explain adaptive behavior by invoking discrete modular processes mediating between stimulation and response. The concomitant modeling strategy of estimating linearly independent effects with normal (Gaussian) patterns of residual variability can be suitable for systems where behaviors result from numerous components. However, they fail to characterize much of the measured behavior, which, rather, exhibit nonlinear correlations across multiple space and time scales entailing non-normal (i.e., non-Gaussian) distributions, such as power-law distributions characterized by skewness. Power-law distributions in empirical measurements often exhibit probability decay or growth characterized by fractional exponents. Systems whose distributions adhere to a single power law are commonly called ``fractal'' and, when conforming to various power laws, ``multifractal.'' Simulated systems, where variations in individual components (i.e., fluctuations) correlate across space and time scales with those of other components within the same system, frequently yield measurements with fractal and multifractal characteristics \cite{lovejoy2018weather, sornette2006critical}. As a result, both fractality and multifractality in empirical time series of biological and psychological measurements have been interpreted as compelling evidence suggesting interdependence \cite{kelty2013tutorial, kelty2021multifractal, ihlen2010interaction, ihlen2012introduction, ihlen2013multifractal, van2003self, van2005human}. Hence, an important new strain of theorizing has proposed to pick up on earlier attempts (e.g., \cite{turing1952chemical}) to characterize the nonlinear fluid dynamical mechanisms governing the growth and behavior of living, thinking systems \cite{jablonkaa2023interacting, nicholson2018everything, west2022fractal}. The present work considers how we might elaborate simulations of cascade dynamical models to begin to elaborate theoretical expectations for the multifractal patternings that we might find empirically if organisms do indeed operate upon these cascade-dynamical principles. 

The multiplicative interactions among component processes across scales entailed by cascades allow reduction of the intricate biological and psychological processes not into multifarious modular components but rather into a series of power-law (a.k.a., ``singular'') relationships, which serve as indicators for the response to constraints or the growth of systems across multiple scales \cite{furmanek2021postural, kelty2021multifractality, mangalam2021hypothetical, mangalam2021visual}. (Notably, the concept of multiplicative random cascades emerged to infuse intermittency into rainfall patterns \cite{lovejoy1985generalized, waymire1985scaling}. These models have successfully replicated the scale-invariant structures present in spatially disaggregated rainfall.) As we expand our theoretical frameworks to encompass the brain, body, and contextual surroundings, we discover these entities are replete with cascades. Even as power-law behavior can deteriorate and manifest diversely, even at specific scales, the generality of power-law forms becomes a versatile language to describe the typical behavior of the brain, body, and context. The term ``universal'' has traditionally been used to describe the power-law behavior of cascades, suggesting that critical dynamics, characterized by power laws, may be prevalent across disparate systems, regardless of their material composition \cite{lovejoy2018weather}. Furthermore, power-law behavior can propagate between neighboring systems, providing insights into how cascade behavior permeates biology and psychology. While the exact conditions for this propagation remain uncharted, this conceptual framework offers a geometric perspective on how these entities might synchronize their activities. Empirical estimates of power-law behavior within the brain \cite{kardan2020distinguishing, kardan2023improvements} and throughout bodily movement can support prediction of human cognitive problem solving \cite{anastas2014executive, dixon2012multifractal, stephen2011strong, stephen2012scaling} and of perceptuomotor responses in various task settings \cite{bell2019non, bloomfield2021perceiving, carver2017multifractal, jacobson2021multifractality, kelty2021multifractal, kelty2023multifractalnonlinearity, mangalam2020multifractal, mangalam2020multiplicative, palatinus2014haptic}. Hence, the present manuscript begins from the premise that multifractal geometries allow us to query the role of multiplicative interactions that enact cascades in coordinating organismal behavior.

\subsection{Interactivity across scales in the coordination of organismal behavior: Ergodicity-breaking problems and multifractal solutions}

Whatever the best generative model might be for organisms and however they might manifest multifractal structure, a crucial starting point is to recognize that organisms are not monoliths but richly textured and task-sensitive ensembles of very many degrees of freedom \cite{bernstein1966co, bernstein1996dexterity}. Organisms thrive on their fluid capacity to gather and release their degrees of freedom with constraints they can build or break according to task demands \cite{gottlieb2007probabilistic, lewontin1982organism, pattee2001physics, pattee2007laws, pattee2012physical, pattee2013epistemic}. While we can model the individual observables as topologically one-dimensional series with potentially multifractal dimensionality, at a certain point, behavioral and biological sciences may profit from situating these individual observables back into the larger systems that they compose, that they participate in, that they belong to, and that they may help to lead. Ultimately, the observables become interesting to theory because of their capacity to absorb the activity of other observables in these larger systems that comprise them. So, when our empirical approach to a distributed system leads us to pick one or another degree of freedom to focus on, we can, by all means, model a single degree of freedom at a time---we can pick one observable from that one degree of freedom. But the only reason we found that degree of freedom interesting in the first place was its participation in a bigger system. The observable is receiving stimulation (e.g., innervations, collisions, suggestions) and, through its own interactions with the task context, producing new behaviors that feedback to the system at large \cite{latash1996bernstein, latash2008synergy, latash2012bliss, latash2020primitives, latash2020synergies, latash2021bernstein, latash2021one, reschechtko2018stability}.

Tackling this fluid capacity of organisms to meet these task contexts is a rich challenge for biological and psychological sciences. For one, the intermittency entailed by adaptive tailoring of behavior to meet changing intentions and changing task contexts has amounted to various forms of ergodicity breaking in our raw measurements. The incapacity of the linear model to address these ergodicity-breaking measurements has long been known in behavioral sciences \cite{bills1935fatigue}, and despite the formal solutions having roots at least as old as behavioral science recognition of the problem, the solutions developed in formalisms originally native to physical disciplines \cite{richardson1926atmospheric} have only lately found their way to behavioral science applications \cite{kelty2014interwoven, van2003self}. In the time it has taken for potential statistical-physical solutions to reach behavioral-science challenges, the ergodicity problem has grown to full bloom across a wide range of scales and of species \cite{cherstvy2013anomalous, cherstvy2014nonergodicity, hamaker2005statistical, jeon2011vivo, li2022non, metzler2014anomalous, mangalam2021point, molenaar2004manifesto, molenaar2008implications, molenaar2009analyzing, molenaar2009new}. Recent theoretical work has proposed both that cascades may offer a potential account of ergodicity-breaking and that we might solve the challenge of modeling the ergodicity-breaking behaviors in evidence by using multifractal descriptors of the cascade processes \cite{kelty2022fractal, kelty2023multifractaldescriptors, mangalam2022ergodic, mangalam2023ergodic}. Multifractal descriptors of ergodicity-breaking behavior are themselves less ergodicity-breaking than the original raw measurement time series. Hence, ergodic multifractal descriptors meet the assumptions of linear causal models, which is at least an analytical convenience.

The analytical convenience might point to deeper conceptual stakes in our attempt to explain organism function. The validity of ergodic multifractal descriptors for linear causal modeling may further suggest that cascade dynamics characterize at least some of the causal relationships knitting together an organism's many degrees of freedom. The present work aims to capitalize on the multifractal availability of ergodic description, and it begins to ask how the multifractal structure of a single observable might bear the imprint of another multifractal process. That is, the fact that we find multifractal structure in any single observable from an organism suggests, minimally, that we might find multifractal structure in several other observables in the same organism---and we certainly do find this empirically in model organisms \cite{kelty2017threading, mangalam2020global, mangalam2020multifractal}. If an organism constitutes an ensemble of multifractal processes, we do not expect the ensemble to be homogeneous but to exhibit a diversity of multifractal patterns. Certainly, observables that neighbor each other or work together may show similar multifractal structures, and observables farther apart or functionally segregable from one another may be free to show greater disparity in multifractal structures. Amidst all this diversity, we might consider how multifractal patterns flow through an organism. Suppose the organism coordinates its degrees of freedom through cascade-like flows (cf. \cite{jablonkaa2023interacting, nicholson2018everything, west2022fractal}). In that case, the organism-wide patterns of how multifractality shifts and spreads through its many degrees of freedom might provide a window to these flows. So, it could be that the ergodic description via multifractal modeling is not just stable enough to work, and it may also be uniquely poised to show how cascade dynamics support the task-dependently intermittent behavior that has proven so intractable to linear approaches \cite{kelty2021multifractal, kelty2021multifractality, kelty2022turing}.

\subsection{Multifractal flows as part of the coordination among organismal degrees of freedom}

What the network science has already begun to model is the point that multifractality is an elegant operationalization of the cascade relationships that might coordinate an organism's distribution of degrees of freedom---a point that appears theoretically \cite{xiao2021deciphering} and empirically \cite{bell2019non, carver2017multifractal, dixon2012multifractal, kelty2014interwoven, kelty2017threading, soberano2015demystifying}. Also, multifractality may entail more about the broader system than a single observable can embody in any empirical behavioral or biological sense. We know that the extended systems behind our observables can have a spatiotemporal structure more globally \cite{goldberger1990chaos}, and we know that fractal structure has even supported the theoretical expectation that we should be able to unpack more global dynamics from the lower-dimensional projection of these dynamics appearing in our single observables \cite{sauer1991embedology}. Presumably, we have not been lucky enough to pick only the observables with multifractal structure; presumably, the multifractal structure is a more generic property of the whole organismic system. The hierarchical structure implicit in any observable is unlikely to belong strictly to that observable but sooner reflects the aggregates wielding that observable. Hence, we reserve a hope that multifractal structure for any single observable should bear traces of the activity of other observables throughout the system. We expect that the multifractal structure at any single observable could thus carry, at the same time, echoes of the broader system that contains it and echoes of the novel contacts that it engages in. Certainly, we ourselves collected a full body motion of human participants during a perceptual task \cite{mangalam2020bodywide, mangalam2020global}. Rudimentary network analyses addressing the entire, body-wide marker set demonstrated rampant and wide-ranging relationships between anatomical parts across many of the pairwise relationships between current multifractality in one observable and later multifractality in another. The empirical evidence suggests that multifractality at any single observable may bear the imprint of multifractality elsewhere.

The major challenge of this empirical modeling of multifractal networks is that we have yet to develop principled expectations. How should multifractality spread? What potentially cascading spatiotemporal constraints might govern that spreading influence? In one sense, the insights available from studying a single observable's multifractality have potentially blossomed beyond the earlier possibilities that, say, more or less multifractality is associated with health \cite{ivanov1999multifractality}. Yes, the nonlinear properties of a single observable may carry traces of a whole body's functioning \cite{mangalam2020bodywide, mangalam2020multiplicative, rahman2021detecting, sechidis2021machine, tsanas2021remote}. But With these concerns in mind, global coherence of organism-wide coordination may be intermittent with the breaking and release of constraints according to task needs. Nonetheless, multifractal models might support predictions about the long-range patterns of such intermittency (e.g., single degrees of freedom in a complex movement system) in which we can begin to model how a single multifractal observable a participating member of an ensemble. One observable might stand alone in a theoretical vacuum, but it might also be a particle modeled as part of a loosely interactive ensemble. Loose interactions amongst observables within the ensemble will allow that. For instance, simulation work investigating single cascade processes has been crucial for understanding the sequential variations of a single human's forefinger pressing a button, for example, to signal their estimation of $1$-second intervals \cite{kuznetsov2011effects, ihlen2013multifractal}. This strategy is central to furthering a cascade-dynamical portrayal of the finger behavior in that task. However, in this example, the pressing of a fingertip is only the most immediate point of contact between the whole organism and the task environment. As we acknowledge that the single button press is only the tip of an organism-wide iceberg, we recognize that the pressing of the fingertip emerges lawfully from a vast network of anticipatory postural adaptations and longer-range postures distributed across the whole movement system \cite{latash1996bernstein, latash2008synergy, latash2012bliss, latash2020primitives, latash2020synergies, latash2021bernstein, latash2021one, reschechtko2018stability}. Presumably, what we see in the single observable is the endogenous flow of coordinating cascades from neighboring or connected observables endogenous to the organism and any exogenous stimulation (e.g., \cite{kelty2023multifractalstimulation}). The foregoing evidence from multifractal estimates from the brain and body could indicate cascades spreading from one observable to the next. That spatiotemporal spreading of cascades through the organism could provide a way to explain these anticipatory adaptations at many scales. But again, the major challenge is that we have no theoretical guide to our expectations of diagnosing these organism-wide flows from individual-observable multifractal results.

\subsection{The present study: Examining the multifractal nonlinearity of multiplicative and additive cascades incorporating multifractal noise}

\subsubsection{Replicating a test for multifractal nonlinearity in additive vs. multiplicative cascades}

The present work aims to develop novel theoretical expectations about cascade dynamics exerting influences from one observable to the next by incorporating multifractal perturbations into cascade simulations. Years of extensive empirical research employing multifractal formalisms have pursued an enduring question: Do the discernible nonlinear connections spanning various scales arise from multiplicative interactions, or are they inherently embedded within the constituent processes? We want to solidify our interpretations of multifractal nonlinearity from the expectation of multiplicative interactions across scales within biological and psychological processes. To address this gap, this study takes a rigorous approach through numerical simulations, dissecting the intricate interplay between multiplicative interactions and the inherent nonlinear characteristics of component processes in the genesis of multifractal structures. As in prior work \cite{mangalam2024multifractalnonlinear}, we evaluate ``multifractal nonlinearity'' as defined by $t_\mathrm{MF}$, i.e., the $t$-statistic comparing an original cascade time series' multifractal spectrum width $\Delta\alpha_\mathrm{Orig}$ to a sample of phase-randomized surrogates time series' multifractal spectrum widths $\Delta\alpha_\mathrm{Surr}$. Throughout, given our previous finding of cascade multiplicativity interacting with the presence of $fGn$ at each generation \cite{mangalam2024multifractalnonlinear}, we expected multiplicativity to exert a strong effect across the board but also to show interactions with specific noise types.

\subsubsection{Introducing effects of different types of multifractal noise on repeated generations of the cascade simulation}

We developed cascade simulations that applied multifractal noise to each generation. That is, these cascade simulations defined each two children cells in each new generation as a function of their parent's cell proportion multiplied by consecutive pairs of values drawn sequentially from multifractal-noise processes. Specifically, we conducted simulations of cascade time series featuring component processes operating at distinct timescales, each characterized by one of four properties: (i) Multifractal nonlinearity---multifractal structure with nonlinear interactions across scales; (ii) multifractal linearity---the same multifractal structure but arising out of purely linear interactions (obtained via the Iterative Amplitude Adjusted Wavelet Transform of multifractal nonlinearity (IAAWT) \cite{keylock2017multifractal, keylock2019hypothesis}; (iii) Phase-randomized linearity---multifractal structure precisely due to nonlinear interactions removed (obtained via the Iterative Amplitude Adjustment Fourier Transform (IAAFT) \cite{ihlen2012introduction, schreiber1996improved}; (iv) shuffled, which randomized phase- and amplitude-spectra---multifractal structure due to both nonlinear and linear interactions removed (obtained via shuffling).

In line with previous findings \cite{mangalam2024multifractal}, we anticipated that multiplicative interactions---as opposed to additivity---would enhance the singularity at peak values and broaden the left-side width. Simultaneously, we expected noise-dependent variations in the right-side width \cite{mangalam2024multifractal}. Importantly, given that all noise processes stemmed from inherently nonlinear, multiplicative-cascading multifractal noise processes, the resulting histograms would consistently deviate from Gaussian distribution, manifesting variations solely in temporal structure. Thus, our primary inquiry revolved around the impact of multifractal noise at each generation of a multiplicative cascade on the ensuing multifractal spectrum. We hypothesized that the sequential heterogeneity introduced by multifractal noise would generally render the multifractal spectra more asymmetric. This asymmetry would extend towards the left side, accentuating heterogeneity in large-sized events. The collaboration between larger, more abrupt changes induced by the non-Gaussian multiplicative cascade dynamics and the temporal heterogeneity in multifractal noise was expected to synergize, allowing heterogeneity to permeate progressively larger fluctuations.

Specific types of multifractal noise raised an open question of how original nonlinear correlations, the multifractal spectrum width, the linear correlations, or the non-Gaussian histogram would moderate the effect of multiplicativity on multifractal spectra and multifractal nonlinearity. For instance, previous work showed that multiplicative cascades with $fGn$ promoted the persistence of right-side width of the multifractal spectrum across generations and the size and statistical significance of $t_\mathrm{MF}$. A likely candidate for replicating this noise effect would have been IAAFT noise, which carries the same linear correlations as multifractal noise with less of the nonlinear correlations and less of the multifractal spectrum width. That said, the non-Gaussianity and symmetricality of the IAAFT surrogate \cite{porta2008temporal, venema2006statistical, venema2006surrogate, zanin2020time} potentially induces a symmetric trend not necessarily present in all $fGn$ processes. Specifically, the symmetricality of an IAAFT with a $1/f$-like decay of spectral power across spectral frequencies might have a symmetric and convex or concave structure across time, for instance, with the greatest spectral power manifesting symmetrically across the entire series length---and non-Gaussianity is likely to exacerbate this transient trend at the beginning and end of the series. IAAWT series have less symmetry across time, but removing the original nonlinear correlations may still induce a trend of large events characteristic of non-Gaussianity with multifractally comparable temporal heterogeneity. The effect of a potential transient led us to incorporate noise types in two different ways: first, from the beginning of a noise process where potential transients reflecting linearization might be strongest, and second, from the center of the noise process where transient-like trends reflecting linearization might be weakest. Again, we approached this sampling effect with exploratory interest, understanding that it might not have any effect. After all, the generations---particularly early in the cascade simulation---are brief, with only a few parent cells and only two children per parent cell. Hence, just as we can expect little signature of multifractal nonlinearity in a very short series \cite{kelty2023multifractaltest}, we might also expect similarly weak to no expression of nonlinear correlations in the cascade process. However, we saw the opportunity to begin investigating whether trends incident to linear correlations could make multifractal more or less asymmetric particularly as the later generations with a larger sequence of children cells absorbed longer sequences of the noise processes. We could be sure that the shuffled multifractal noise would have the same non-Gaussian histogram but no sequence and, hence, no transients. So, we expected that multiplicative cascades with shuffled multifractal noise could have no such sampling effects on multifractal spectra or multifractal nonlinearity. 

Besides answering the more local biological and psychological sciences question, the present work will also connect with the broader statistical-physical questions of ergodicity breaking. We can be specific to the statistical properties breaking ergodicity (e.g., \cite{lebowitz1973modern}): multiplicative cascades can break ergodicity due to non-Gaussianity, while additive cascades might be more likely to break ergodicity due to correlated sequence. Furthermore, multiplicative cascades may be more sensitive than additive cascades to the sequential ergodicity breaking of constituent noise processes \cite{mangalam2024multifractalnonlinear}. So, as we build cascades with progressively more heterogeneous temporal structure, we might expect that multiplicative cascades would be more likely to show sequence-driven ergodicity breaking, for instance, for originally multifractal-nonlinear- and IAAWT-noise cases as opposed to the IAAFT- and shuffled-noise cases. 

We evaluated three specific hypotheses, considering the range of data collection and processing decisions a behavioral scientist might encounter with multifractal analysis: 

\textbf{Hypothesis 1:} We postulated that multiplicative interactions across scales would amplify the right-side width, left-side width, and peak $\alpha(q=0)$ of cascade time series more significantly than additivity. This amplification translates to heightened temporal heterogeneity for small-sized fluctuations, increased temporal heterogeneity for large-sized fluctuations, and more robust scaling of proportions on average (cf. \cite{schertzer1987physical}). We expected this effect of multiplicative interactions to be stronger for cascades incorporated multifractal-nonlinear noise than for cascades incorporating IAAWT- and IAAFT-noise types sampled from the beginning but not stronger than for cascades incorporating IAAWT- and IAAFT-noise types sampled from the center of the noise process after any linear transients. 

\textbf{Hypothesis 2:} We hypothesized that multiplicativity would amplify the emergence of multifractality as the multiplicative cascade unfolds across successive generations (cf. \cite{passos2011fat}). We expected this effect to transcend the influence of the growing length of the cascade time series implicit in the progressive generations of a traditional binomial multiplicative cascade model. We expected this effect to be stronger for cascades incorporating multifractal-nonlinear noise than for cascades incorporating IAAWT- and IAAFT-noise types sampled from the beginning but not stronger than for cascades incorporating IAAWT- and IAAFT-noise types sampled from the center of the noise process after any linear transients.

\textbf{Hypothesis 3:} We hypothesized that additive and multiplicative cascades would each break ergodicity due to different statistical properties as we had found previously, that is, due to non-Gaussianity and correlated sequence (i.e., as compared to shuffled surrogates). As in the foregoing hypotheses, we expected this effect of multiplicativity to be stronger for the multifractal-nonlinear- and IAAWT-noise cases than for the IAAFT- and shuffled-noise cases.

\section{Methods}

\subsection{Generating cascade time series}

\subsubsection{Binomial fracturing and binomial noise terms in prior cascade simulation}

\begin{figure*}
    \includegraphics[width=6.125in]{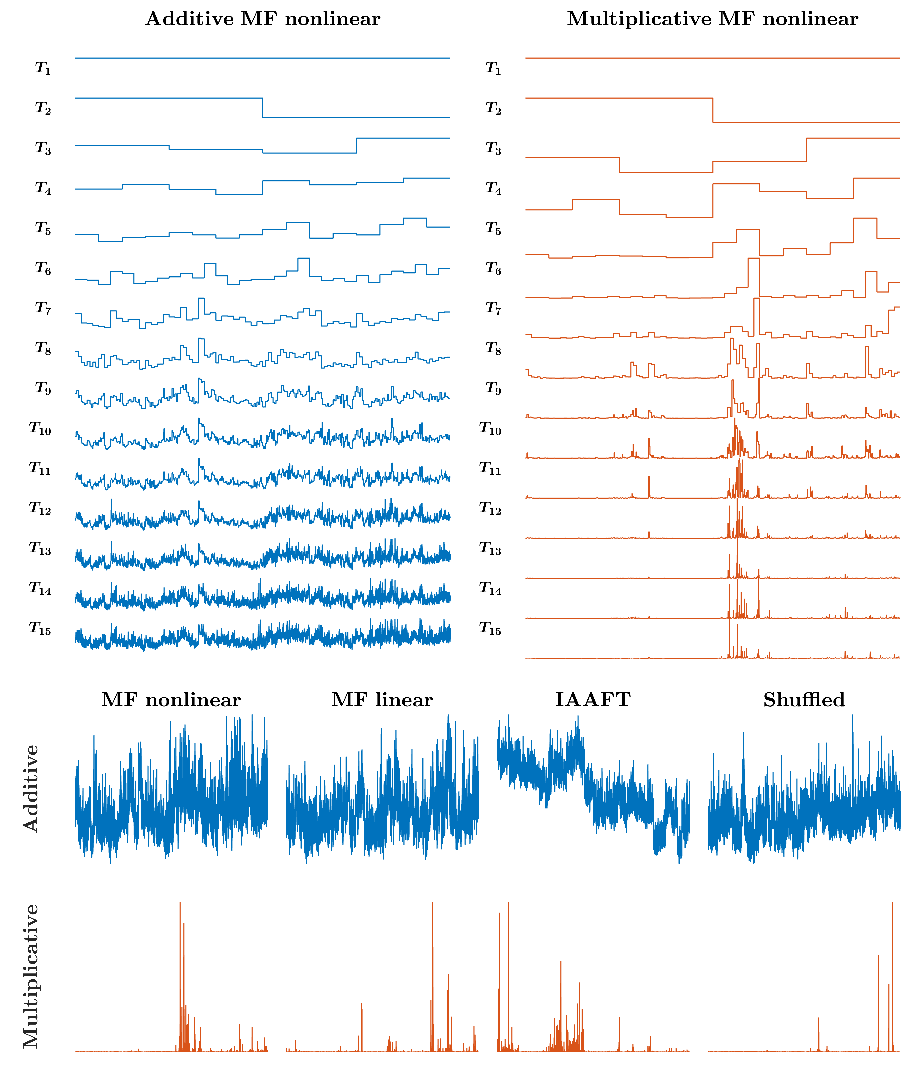}
    \caption{A binomial additive or multiplicative cascade serves as a mathematical framework for elucidating how the distribution of quantities or events can evolve across increasingly smaller sample sizes and shorter time intervals. In the \textit{upper-left} quadrant, we observe an additive cascade characterized by multifractal (MF) nonlinear noise terms manifesting across $15$ generations of iterative splitting and additive interactions. Meanwhile, in the \textit{upper-right} quadrant, a multiplicative cascade with MF nonlinear noise terms unfolds over $15$ generations, showcasing multiplicative interactions. Each curve has been normalized by its maximum value and vertically shifted by $1$ unit to enhance clarity. The \textit{lower panel} presents a compelling display of both cascade types, exemplifying variations in interactivity that span from additive to multiplicative. Both cascade types encompass various noise types, including nonlinear multifractal noise, IAAWT-specified linear multifractal noise retaining both amplitude spectrum and multifractal spectrum width, IAAFT-specific linear noise maintaining only the amplitude spectrum, and random shuffling of the originally nonlinear multifractal noise. Observe the pronounced prevalence of exceedingly large events in multiplicative cascades, not additive cascades.}
    \label{fig: f1}
\end{figure*}

Cascades involve an iterative process spanning $g$ generations, where $n_{g}$ cells in each generation manipulate the proportions $p_{i,j}$ contained within them. Here, $i$ ranges from $1$ to $n_{j}$ for generation $j$ less than $g$. In the subsequent generation ($j+1$), there are $n_{c}$ children cells, each inheriting proportions denoted as $p_{k,j+1}$, where $i$ ranges from $1$ to $n_{j+1}$. In binomial cascades, each parent cell distributes proportions to two children cells in the next generation (i.e., $n_{c}=2$). As elaborated below, we generated eight distinct types of binomial cascades, half following a multiplicative pattern while the other half adhering to an additive pattern. In multiplicative cascades, the proportions allocated to the $n_{c}$ children cells result from $n_{c}$ distinct multiplicative operations governing the distribution of the parent cell's proportion. Conversely, in additive cascades, the proportions are determined by $n_{c}$ distinct addition operations. Our cascades underwent binomial fracturing at each generation, with binomial noise terms conventionally employed to determine the pairings of children cells (Fig.~\ref{fig: f1}). Applying these binomial noise terms could follow a deterministic approach for any $i$th parent cell in the $j$th generation, for instance, the same noise terms, $W_{1}=0.25$ and $W_{2}=0.75$, could be applied consistently to calculate the proportions in the $2i-1$th and the $2i$th children cells in the $j+1$th generation, resulting in $p_{2i-1,j+1}=p_{i,j}\cdot W_{1}$ and $p_{2i,j+1}=p_{i,j}\cdot W_{2}$, respectively.

\subsubsection{Beyond binomial noise terms: Binomial-fracturing cascades with noise terms defined across entire generations to test for effects of length, $fGn$, and multiplicativity}

In this study, our primary objective revolves around preserving the binomial fracturing pattern from parent to offspring cells while introducing a random element to the noise terms, extending beyond the confines of the binomial ($W(1)$,$W(2)$) structure. The pivotal deviation in our approach involves generating the cascade time series by introducing noise terms spanning the entire generation. These noise terms exhibited variations in interactivity, ranging from additive to multiplicative, and encompass diverse noise types, including multifractally nonlinear, multifractally linear, phase-randomized, and phase- and amplitude-randomized variations. We can now precisely delineate the eight cascade types we simulated using these terms.

\textbf{\textit{Multifractally nonlinear noise.}} We generated multifractally nonlinear noise using an established cascade model \cite{kiyono2007estimator} that shows the deformation predicted by the log-normal model of Kolmogorov \cite{kolmogorov1962refinement} and Obukhov \cite{obukhov1962some}. Our numerical procedure for generating a time series from our model unfolds as follows. To initiate the process, we generate a time series denoted as ${\xi_{i}}$, comprised of Gaussian white noise with a mean of zero. This series's total number of data points equals $2^{m}$, with $i$ ranging from $1$ to $2^{m}$, where $m$ represents the total number of cascade steps. In the initial cascade step, denoted as $j=1$ ($j<m$), we partition the entire interval into two equal subintervals. Subsequently, we multiply each element $\xi_{i}$ within these subintervals by random weights defined as $\exp{[\omega^{(1)}(k)]}$, where $k$ takes on values of $0$ and $1$. Notably, the $\omega^{(1)}(k)$ values constitute independent Gaussian random variables, each possessing a mean of zero and a consistent variance of $\langle \omega^{(1)}(k)^{2} \rangle = \lambda_{0}^{2}$. As we progress to the subsequent cascade step, designated as $j=2$, we subdivide each existing subinterval into two equal segments. Here, we apply the same random weights $\exp{[\omega^{(1)}(k)]}$, with $k$ spanning values from $0$ to $3$. This process continues iteratively, preserving consistency until we have completed the $m$ cascade steps. The resulting time series, denoted as ${x_{i}}$, represents the outcome of this cascade process and is expressed as:
\begin{equation*}
    x_{i}=\xi_{i}\,\exp{\Bigg[\sum_{j=1}^{m}\omega^{(j)}\bigg(\bigg\lfloor\frac{i-1}{2^{m-j}}\bigg\rfloor\bigg)\Bigg]}, \tag{1}\label{eq: 1}
\end{equation*}
where $\lfloor \cdot \rfloor$ is the floor function. The random variable $x_{i}$ is not standardized to simplify the notation. We generated such multifractally nonlinear noise of length $l=2^{15}$ (i.e., $m=15$) with its probability density function (PDF) described by $\lambda^{2}=m\lambda_{0}^{2}$ according to the Castaing's equation \cite{castaing1990velocity}; we used $\lambda^{2}=0.5$. We used these to obtain the other three noise types, which ultimately were used to generate the random additive and multiplicative cascades subjected to multifractal and ergodicity-breaking analysis to test the above hypotheses.

\textbf{\textit{Multifractally linear noise.}} To generate a multifractally linear version of this multifractally nonlinear noise series, we used the Iterative Amplitude Adjusted Wavelet Transform (IAAWT) method, as outlined in \cite{keylock2017multifractal, keylock2019hypothesis}. In the context of a time series with a length of $l=2^{j}$, the IAAWT procedure unfolds as follows:
\begin{itemize}
    \item Execute a dual-tree complex discrete wavelet transform, extracting both amplitudes and phases across all $j$ scales for the complex-valued $w_{j,k}$ with $k$ ranging from $1$ to $2^{j-1}$ at each $j$;
    \item Randomly rearrange the original data and employ the dual-tree complex DWT to generate randomized wavelet phases at each scale.;
    \item Generate updated $w_{j,k}$ by merging the initial amplitudes with the randomized phases.
    \item Continuously iterate through the subsequent steps until convergence is attained:
    \begin{itemize}
        \item Execute the inverse wavelet transform to generate a fresh time series, followed by applying the identical amplitude adjustment step in the IAAFT algorithm.
        \item Utilize the dual-tree complex DWT to extract the updated phases and then amalgamate these with the initial amplitudes to produce the most recent $w_{j,k}$.
  \end{itemize}
\end{itemize}
The IAAWT method retains the original dataset's probability distribution while preserving its multifractal structure up to a stringent convergence criterion. We generated ten IAAWT series for each original noise series and identified the one that exhibited the closest resemblance to the original noise series in multifractal spectrum width.

\textbf{\textit{Phase-randomized, IAAFT noise.}} To generate a phase-randomized version of this multifractally nonlinear noise series, we used the Iterative Amplitude Adjusted Fourier Transform (IAAFT) method, as outlined in \cite{ihlen2012introduction, schertzer1987physical}. In the context of a time series $x_{t},t=1,2,\dots,N$, the IAAFT procedure unfolds as follows:
\begin{itemize}
    \item Capture and retain the squared amplitudes derived from the discrete Fourier transform of $x_{t}$ (i.e., $X_{f}^{2}=|\sum_{1}^{N}x_{t}\,e^{i2\pi\,f(t/N)}|^{2}$);
    \item Initiate the process by randomly shuffling $x_{t}$ to yield $x_{t}^{(j=0)}$;
    \item Proceed by iteratively executing a power spectrum step followed by a rank-order matching step on $x_{t}^{(j)}$ in the following manner:
    \begin{itemize}
        \item Perform a Fourier transformation on $x_{t}^{(j)}$ and update the squared amplitudes with $X_{f}^{2}$, keeping the phases intact. Given the initial random sorting, this process should conserve the spectrum, albeit with randomized phases. Subsequently, invert the Fourier transformation, restoring the original amplitudes;
        \item Substitute the values within the fresh series $x_{t}^{(j)}$ with those from $x_{t}$ through a rank-order matching procedure. While this approach maintains the integrity of the original dataset, it does compromise the precision of spectral matching, thereby elucidating the approximate replication of Fourier amplitudes;
   \end{itemize}
   \item Continue the process until either the convergence criterion is satisfactorily met or alterations become negligible, thus rendering any re-ordering of values from the prior iteration unnecessary.
\end{itemize}
In this way, the histogram of the original data is preserved precisely, and the Fourier spectrum is approximated to a given error tolerance.

\textbf{\textit{Phase- and amplitude-randomized, shuffled noise.}} We subjected this multifractally nonlinear noise series to $100$ shuffles, resulting in a version characterized by randomized phases and amplitudes. Shuffling effectively erases both the phase and amplitude spectra of the series, thereby nullifying any characteristics associated with the original temporal sequence of values.

Finally, we used these four kinds of noise series to generate the following eight cascade types:
\begin{enumerate}
    \item \textbf{\textit{Additive multifractal nonlinear}}. Noise terms for generation $j+1$ include $x_{2^{j+1}},x_{2^{j+1}+1},\dots,x_{2^{j+2}-1}$ from multifractally nonlinear noise, and the $2i-1$th and $2i$th children cells in the $j+1$th generation holds proportion $p_{i}+W_{2i-1,j+1}$ and $p_{i}+W_{2i,j+1}$, respectively.
    \item \textbf{\textit{Additive multifractal linear}}. Noise terms for generation $j+1$ include $x_{2^{j+1}},x_{2^{j+1}+1},\dots,x_{2^{j+2}-1}$ from multifractally linear noise, and the $2i-1$th and $2i$th children cells in the $j+1$th generation holds proportion $p_{i}+W_{2i-1,j+1}$ and $p_{i}+W_{2i,j+1}$, respectively.
    \item \textbf{\textit{Additive IAAFT.}} Noise terms for generation $j+1$ include $x_{2^{j+1}},x_{2^{j+1}+1},\dots,x_{2^{j+2}-1}$ from phase-randomized IAAFT noise, and the $2i-1$th and $2i$th children cells in the $j+1$th generation holds proportion $p_{i}+W_{2i-1,j+1}$ and $p_{i}+W_{2i,j+1}$, respectively.
    \item \textbf{\textit{Additive shuffled.}} Noise terms for generation $j+1$ include $x_{2^{j+1}},x_{2^{j+1}+1},\dots,x_{2^{j+2}-1}$ from phase- and amplitude-randomized, shuffled, and the $2i-1$th and $2i$th children cells in the $j+1$th generation holds proportion $p_{i}+W_{2i-1,j+1}$ and $p_{i}+W_{2i,j+1}$, respectively.
    \item \textbf{\textit{Multiplicative multifractal nonlinear}}. Noise terms for generation $j+1$ include $x_{2^{j+1}},x_{2^{j+1}+1},\dots,x_{2^{j+2}-1}$ from multifractally nonlinear noise, and the $2i-1$th and $2i$th children cells in the $j+1$th generation holds proportion $p_{i}\cdot W_{2i-1,j+1}$ and $p_{i}\cdot W_{2i,j+1}$, respectively.
    \item \textbf{\textit{Multiplicative multifractal linear}}. Noise terms for generation $j+1$ include $x_{2^{j+1}},x_{2^{j+1}+1},\dots,x_{2^{j+2}-1}$ from multifractally linear noise, and the $2i-1$th and $2i$th children cells in the $j+1$th generation holds proportion $p_{i}\cdot W_{2i-1,j+1}$ and $p_{i}\cdot W_{2i,j+1}$, respectively.
    \item \textbf{\textit{Multiplicative IAAFT.}} Noise terms for generation $j+1$ include $x_{2^{j+1}},x_{2^{j+1}+1},\dots,x_{2^{j+2}-1}$ from phase-randomized IAAFT noise, and the $2i-1$th and $2i$th children cells in the $j+1$th generation holds proportion $p_{i}\cdot W_{2i-1,j+1}$ and $p_{i}\cdot W_{2i,j+1}$, respectively.
    \item \textbf{\textit{Multiplicative shuffled.}} Noise terms for generation $j+1$ include $x_{2^{j+1}},x_{2^{j+1}+1},\dots,x_{2^{j+2}-1}$ from phase- and amplitude-randomized, shuffled, and the $2i-1$th and $2i$th children cells in the $j+1$th generation holds proportion $p_{i}\cdot W_{2i-1,j+1}$ and $p_{i}\cdot W_{2i,j+1}$, respectively.
\end{enumerate}
To shed light on the underlying factors fueling multifractal nonlinearity within biological and psychological processes, we constructed cascades comprising $2^{14}$ samples in their $15$th and concluding generation. This choice of sample length mirrors the typical empirical time series duration commonly encountered in these domains. To overcome any effects of stochasticity in the generation of these cascades and to ensure robustness in our analysis, we simulated a total of $100$ cascade time series for each type.

\subsubsection{Padding cascades with consecutive repetitions of cell values to disentangle generation number from length}

Hypothesis 2 suggests that the number of successive generations plays a pivotal role in these phenomena. To disentangle the effects of length from those of the number of generations, we extended each generation within the original cascades from the $9$th to the $15$th generation by padding them with repeated values, ensuring they all shared a uniform length of $2^{14}$. We applied a similar technique to the surrogate series to ensure a uniform $2^{14}$ length. Notably, we only enforced this specific sequence length for the $9$th through the $15$th generations, as we aimed to maintain the highest confidence in the reliability of the IAAFT surrogates for the original series, particularly for cases where $l\leq2^{9}$.

\subsection{Multifractal and ergodicity-breaking analysis}

We computed key metrics for each simulated series across generations $9$ through $15$ to gain insights into its multifractal characteristics. These metrics encompassed the multifractal spectrum width, denoted as $\Delta\alpha$, which we derived employing Chhabra and Jensen's direct method \cite{chhabra1989direct}. We also delved into the $t$-statistic, symbolized as $t_\mathrm{MF}$, serving as a yardstick for contrasting the multifractal spectrum width $\Delta\alpha_\mathrm{Orig}$ of the original series against a set of surrogate samples characterized by their respective multifractal spectrum widths $\Delta\alpha_\mathrm{Surr}$. In tandem with these assessments, we assessed ergodicity breaking, employing the Thirumalai-Mountain metric ($EB$) \cite{rytov1989principles, thirumalai1989ergodic}. This metric quantified ergodicity breaking in the original series and extended its reach to randomized shufflings of the same series. It is important to note that while calculating $t_\mathrm{MF}$ leveraged phase-randomized surrogates to gauge the original series, the evaluation of $EB$ necessitated comparisons with randomized shufflings. This deliberate choice arises because shuffling perturbs linear and nonlinear correlational sources, contributing to breaking ergodicity.

\subsubsection{Multifractal analysis}

We used Chhabra and Jensen’s \cite{chhabra1989direct} direct method for all analyses. This method estimates multifractal spectrum width $\Delta\alpha$ by sampling a series $x(t)$ at progressively larger scales using the proportion of signal $P_{i}(n)$ falling within the $v$th bin of scale $n$ as
\begin{equation*}
    P_{v}(n)=\frac{\sum\limits_{k=(v-1)\,n+1}^{v\cdot N_{n}}x(k)}{\sum{x(t)}},\quad n=\{4,8,16,\dots\}<T/8. \tag{2}\label{eq: 2}
\end{equation*}
As $n$ increases, $P_{v}(n)$ represents a progressively larger proportion of $x(t)$,
\begin{equation*}
    P(n)\propto n^{\alpha}, \tag{3}\label{eq:3}
\end{equation*}
suggesting a growth of the proportion according to one ``singularity'' strength $\alpha$ \cite{mandelbrot1982fractal}. $P(n)$ exhibits multifractal dynamics when it grows heterogeneously across time scales $n$ according to multiple singularity strengths, such that
\begin{equation*}
    P(n_{v})\propto n^{\alpha_{v}}, \tag{4}\label{eq: 4}
\end{equation*}
whereby each $v$th bin may show a distinct relationship of $P(n)$ with $n$. The width of this singularity spectrum, $\Delta\alpha=(\alpha_{max}-\alpha_{min})$, indicates the heterogeneity of these relationships \cite{halsey1986fractal,mandelbrot2013fractals}.

Chhabra and Jensen's \cite{chhabra1989direct} method estimates $P(n)$ for $N_{n}$ nonoverlapping bins of $n$-sizes and transforms them into a ``mass'' $\mu(q)$ using a $q$ parameter emphasizing higher or lower $P(n)$ for $q>1$ and $q<1$, respectively, in the form
\begin{equation*}
    \mu_{v}(q,n)=\frac{\bigl[P_{v}(n)\bigl]^{q}}{\sum\limits_{j=1}^{N_{n}}\bigl[ P_{j}(n)\bigl]^{q}}. \tag{5}\label{eq: 5}
\end{equation*}
Then, $\alpha(q)$ is the singularity for mass $\mu$-weighted $P(n)$ estimated as
\begin{equation*}
    \alpha(q)=-\lim_{N_{n}\to\infty}\frac{1}{\ln{N_{n}}}\sum_{v=1}^{N_{n}}\mu_{v} (q,n)\ln P_{v}(n)
\end{equation*}
\begin{equation*}
    = \lim_{n\to 0}\frac{1}{\ln{n}}\sum_{v=1}^{N_{n}}\mu_{v}(q,n)\ln{P_{v}(n)}. \tag{6}\label{eq: 6}
\end{equation*}
Each estimated value of $\alpha(q)$ belongs to the multifractal spectrum only when the Shannon entropy of $\mu(q,n)$ scales with $n$ according to the Hausdorff dimension $f(q)$ \cite{chhabra1989direct}, where
\begin{equation*}
    f(q)=-\lim_{N_{n}\to\infty}\frac{1}{\ln N_{n}}\sum_{v=1}^{N_{n}}\mu_{v}(q,n)\ln\mu_{v}(q,n)
\end{equation*}
\begin{equation*}
    = \lim_{v\to0}\frac{1}{\ln{n}}\sum_{v=1}^{N_{n}}\mu_{v}(q,n)\ln{\mu_{v}(q,n)}. \tag{7}\label{eq: 7}
\end{equation*}

For values of $q$ yielding a strong relationship between Eqs.~(\ref{eq: 6}) \& (\ref{eq: 7})---in this study, correlation coefficient $r>0.95$, the parametric curve ${\alpha(q),f(q)}$ or $(\alpha,f(\alpha))$ constitutes the multifractal spectrum and $\Delta\alpha$ (i.e., $\alpha_{max}-\alpha_{min}$) constitutes the multifractal spectrum width. $r$ determines that only scaling relationships of comparable strength can support the estimation of the multifractal spectrum, whether generated as cascades or surrogates. Using a correlation benchmark aims to operationalize previously raised concerns about mis-specifications of the multifractal spectrum \cite{zamir2003critique}.

Our next objective was to discern whether a nonzero $\Delta\alpha$ truly signified multifractality arising from intricate nonlinear interactions across various timescales. To achieve this objective, we compared $\Delta\alpha$ values between the original series and $32$ IAAFT surrogates \cite{ihlen2012introduction, schreiber1996improved} for each simulated series across generations $9$ through $15$. IAAFT stands out as a method capable of symmetrically reshuffling the original values around their autoregressive structure. Consequently, it generates surrogates that disentangle the phase ordering of spectral amplitudes within the series while preserving the linear temporal correlations. The one-sample $\textit{t}$-statistic, $\textit{t}_\mathrm{MF}$, comes into play by computing the difference between $\Delta\alpha$ for the original series and the corresponding values for the $32$ surrogates, which is then divided by the standard error of the spectrum width for these surrogates, facilitating a robust statistical assessment of multifractal nonlinearity.

\subsubsection{Estimating ergodicity breaking parameter for cascade time series}

The dimensionless statistic $EB$, the Thirumalai-Mountain metric can quantify the degree to which a series breaks ergodicity \cite{he2008random, thirumalai1989ergodic, rytov1989principles} as the variance of sample variance divided by the total-sample squared variance:
\begin{equation*}
    EB(x(t))=\frac{\Bigl\langle\Bigl[\overline{\delta^{2}(x(t))}\Bigl]^{2}\Bigl\rangle-\Bigl\langle\overline{\delta^{2}(x(t))}\Bigl\rangle^{2}}{\Bigl\langle\overline{\delta^{2}(x(t))}\Bigl\rangle^{2}}, \tag{8}\label{eq: 8}
\end{equation*}
where $\overline{\delta^{2}(x(t))}=\int_{0}^{t-\Delta}[x(t^{\prime}+\Delta)-x(t^{\prime})]^{2}dt^{\prime}\bigl/(t-\Delta)$ is the time average mean-squared displacement of the stochastic series $x(t)$ for lag time $\Delta$. Ergodicity with least breaking appears as rapid decay of $EB$ to $0$ for progressively larger samples, that is, $EB\rightarrow0$ as $t\rightarrow \infty$. Thus, for Brownian motion $EB(x(t))=\frac{4}{3}(\frac{\Delta}{t})$ \cite{cherstvy2013anomalous, metzler2014anomalous}. Slower decay indicates progressively more ergodicity breaking as in systems with less reproducible or representative trajectories, and no decay or convergence to a finite asymptotic value indicates strong ergodicity breaking \cite{deng2009ergodic}. $EB(x(t))$ thus allows for estimating how much a given time series fulfills ergodic assumptions or breaks ergodicity. Despite the traditional convention of respecting ergodicity as a dichotomy, $EB$ offers a window on how continuous processes can exhibit gradually more or less breaking of ergodicity \cite{kelty2022fractal, kelty2023multifractaldescriptors, mangalam2021point, mangalam2022ergodic, mangalam2023ergodic}. For instance, Deng and Barkai \cite{deng2009ergodic} have shown that for $fBm$,
\begin{multline*}
    EB(x(t))=\\ 
    \begin{cases}
        k(H_\mathrm{fGn})\frac{\Delta}{t}&0<H_\mathrm{fGn}<\frac{3}{4}\\
        k(H_\mathrm{fGn})\frac{\Delta}{t}\ln{t}&H_\mathrm{fGn}=\frac{3}{4}\\
        k(H_\mathrm{fGn})(\frac{\Delta}{t})^{4-4H_\mathrm{fGn}}&\frac{3}{4}<H_\mathrm{fGn}<1.
    \end{cases}
    \tag{9}\label{eq: 9}
\end{multline*}

We computed $EB$ for the original and a shuffled version of each cascade time series across generations $9$ through $15$ (range $=T/8$; lag $\Delta=2$ samples). The shuffled version allows us to distinguish between two types of ergodicity breaking. Indeed, recent theorizing about ergodicity has elaborated the concept beyond simply the stability of a time-varying process around an ensemble average \cite{lebowitz1973modern}. This recent theorizing has distinguished that a process can fail to be ergodic if its temporal sequence exhibits heterogeneity even with the stability of time averages around an ensemble average. Hence, whereas traditional ergodicity breaking involves the failure of Gaussian-like stability, ergodicity breaking can also manifest through sequence. The overall height of the $EB$-vs.-$t$ relationship indicates ergodicity breaking due to non-Gaussian histograms \cite{kelty2022fractal}, and the difference between the decay of the original $EB$-vs.-$t$ curve compared to that for its surrogates indicates ergodicity breaking dependent on sequence \cite{deng2009ergodic, mangalam2022ergodic}.

\section{Results}

\begin{figure*}
    \includegraphics[width=5.5in]{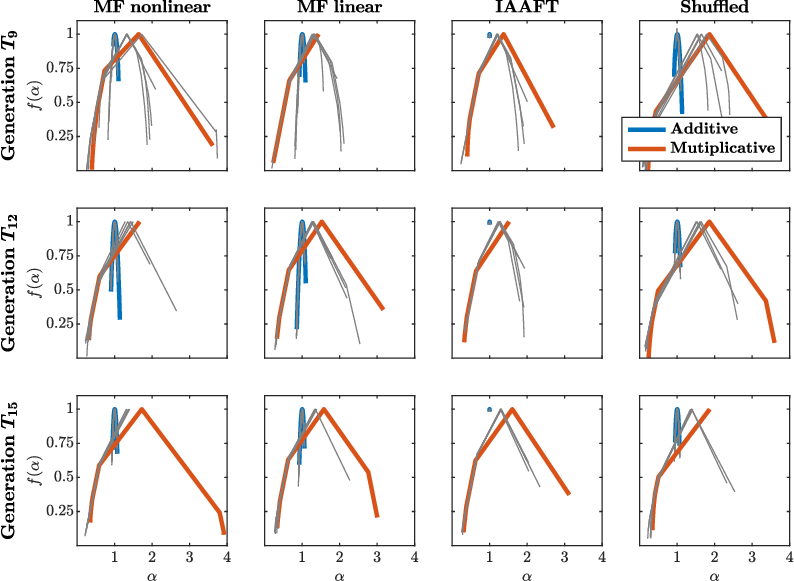}
    \caption{Representative multifractal spectrum for the four additive and multiplicative cascade types at the $9$th, $12$th, and $15$th generations (\textit{top}, \textit{middle}, and \textit{bottom}, respectively) involving the application of originally nonlinear multifractal noise, IAAWT-specified linear multifractal noise retaining both amplitude spectrum and multifractal spectrum width, IAAFT-specific linear noise maintaining only the amplitude spectrum, and random shuffling of the originally nonlinear multifractal noise (in the far-left, center-left, center-right, far-right columns, respectively). The colored lines represent the original spectra, while the grey lines depict five representative surrogate spectra generated using the IAAFT method for each original spectrum. Readers may find it challenging to identify the spectra corresponding to the additive cascades with IAAFT-specified noise; these spectra are so narrow as to appear at this grain as dots located at roughly ($\alpha=1,f(\alpha)=1$) where the other all-wider additive cascades have their peaks.}
    \label{fig: f2}
\end{figure*}

\begin{figure*}
    \includegraphics[width=5.25in]{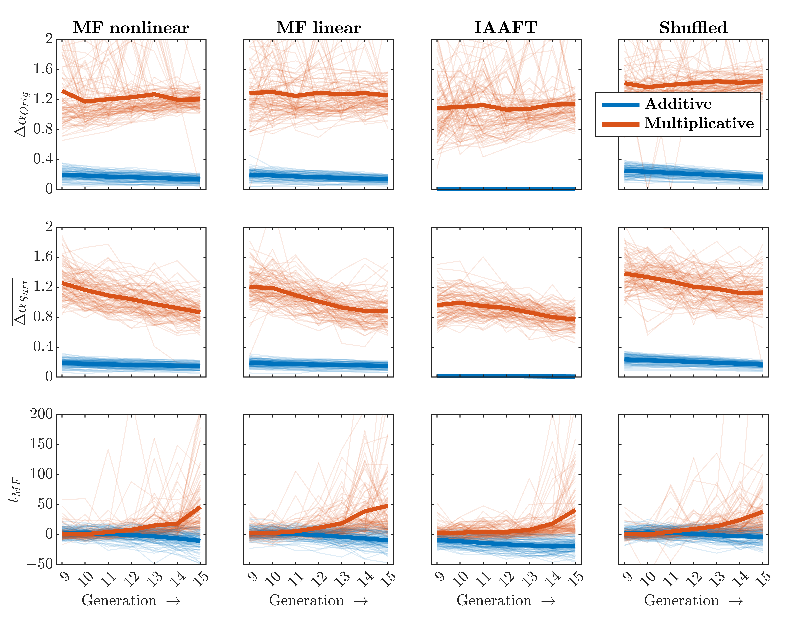}
    \caption{Multifractal spectral properties of the four additive and multiplicative cascade types across successive generations. Multifractal spectrum width for the original cascades $\Delta\alpha_\mathrm{Orig}$ (\textit{top}), multifractal spectrum width for the corresponding $32$ IAAFT surrogates $\overline{\Delta\alpha_\mathrm{Surr}}$ (\textit{middle}), and multifractal nonlinearity $t_\mathrm{MF}$ (\textit{bottom}) involving the application of originally nonlinear multifractal noise, IAAWT-specified linear multifractal noise retaining both amplitude spectrum and multifractal spectrum, IAAFT-specific linear noise maintaining only the amplitude spectrum, and random shuffling of the originally nonlinear multifractal noise (\textit{far-left}, \textit{center-left}, \textit{center-right}, \textit{far-right}, respectively). The bold lines represent the average values derived from $N=100$ numerical simulations, while the finer lines alongside them portray individual data points within these simulation sets.}
    \label{fig: f3}
\end{figure*}

We examined how the two properties---multifractality and multifractal nonlinearity---grow across successive generations of binomial additive and multiplicative cascades to investigate the relationship between multifractality, multifractal nonlinearity, and ergodicity breaking within an additive or a multiplicative cascade. We defer reviewing the ergodicity-breaking results until we describe outcomes related to Hypothesis 2. 

\subsection{Hypothesis 1: Multiplicative and not additive interactions dictate multifractal emergent properties}

For now, we detail initial observations of how these multifractal spectra differ across the different types of cascade simulations. Fig.~\ref{fig: f2} shows an example of a multifractal spectrum for the additive and multiplicative variants of cascades applying each of the five noise types at three generations. We ran a linear mixed-effect regression model of three different multifractal spectral attributes to elaborate beyond these cursory portrayals. Our regression model included five covariates:
\begin{enumerate}
    \item \textbf{\textit{Feature,}} encoding three types of multifractal-analytical outcomes \cite{morales2002wavelet}, namely, a baseline value of $\alpha_{max}-\alpha(q=0)$ (i.e., spectral half-width to the \textit{right} of the peak) and two alternative values of $\alpha(q=0)$ (i.e., the location of the spectral peak) and $\alpha(q=0)-\alpha_{min}(q)$ (spectral half-width to the \textit{left} of the peak). 
    \item \textbf{\textit{Type,}} encoding the four types of cascade as described above indicating the kind of multifractal noise, with a baseline value of \textit{nonlinear} and three alternative values \textit{linear}, \textit{IAAFT}, \textit{shuffled}. 
    \item \textbf{\textit{Multiplicativity,}} encoding the way that cascade simulations applied noise across generations, with a baseline value of \textit{additive} and alternative value \textit{multiplicative}.
    \item \textbf{\textit{Generation,}} encoding the number of generations from $9$ through $15$.
    \item \textbf{\textit{Center,}} encoding whether the cascade sampled from the left of the noise time series ($Center = 0$) or towards the center of the noise time series ($Center = 1$)
\end{enumerate}
We used a full-factorial regression model, including the highest-order interaction covariate Feature $\times$ Type $\times$ Multiplicative $\times$ Generation $\times$ Center and all component lower-order interactions and main effects. We used the function \texttt{lmer} from the package \texttt{lme4} \cite{bates2009package} in $R$ \cite{team2013r}. We detail the outcomes for each of the three features in the following paragraphs before detailing the outcome for each of our hypotheses.

The right-side width of the multifractal spectrum, $\alpha_{max}-\alpha(q=0)$, grew larger by the $9$th generation with multiplicativity and narrowed with subsequent generations. Additive cascades with \textit{multifractal nonlinear} noise had nonzero right-side width ($B=1.03\times10^{-1}, SE=1.23\times10^{-2}, P<0.0001$). Additive cascades with \textit{IAAFT} noise reduced this right-side width almost to zero ($B=-9.68\times10^{-2}, SE=1.73\times10^{-2}, P<0.0001$), but no other noise had any effect in additive cascades. Multiplicative cascades with \textit{multifractal nonlinear} noise nearly doubled the right-side width at $9$th generation ($B=8.26\times10^{-2}, SE=1.73\times10^{-2}, P<0.0001$) but also predicted almost triple the \textit{nonlinear} noise's rate of narrowing the right-side width with subsequent generations ($B=-1.51\times10^{-2}, SE=4.26\times10^{-3}, P=0.0004$). With \textit{multifractal linear} and \textit{IAAFT} noises, the effect of multiplicativity on $9$th-generation right-side width doubled and almost quadrupled, respectively ($B\mathrm{s}=1.22\times10^{-1}\;\mathrm{and}\;3.00\times10^{-1}, SE\mathrm{s}=2.45\times10^{-2}\;\mathrm{and}\;2.45\times10^{-2}, P\mathrm{s}<0.0001$, respectively). Similarly, \textit{multifractal linear} and \textit{IAAFT} noises also doubled and quadrupled, respectively, the multiplicative cascades' narrowing of the right-side width with generations ($B\mathrm{s}=-1.89\times10^{-2}\;\mathrm{and}\;-5.17\times10^{-2}, SE\mathrm{s}=6.03\times10^{-3}\;\mathrm{and}\;6.03\times10^{-3}, P\mathrm{s}=0.002\;\mathrm{and}\;<0.0001$, respectively). However, sampling the noise processes from the center appears to have canceled many effects of these linear noises. Center-sampled \textit{IAAFT} noise canceled the right-side width diminution for additive cascades ($B=1.08\times10^{-1}, SE=2.45\times10^{-2}, P<0.0001$). Center-sampled \textit{multifractal linear} and \textit{IAAFT} noises canceled out the growth of the right-side width on the $9$th-generation's right-side width ($B\mathrm{s}=-1.66\times10^{-1}\;\mathrm{and}\;-3.28\times10^{-1}, SE\mathrm{s}=3.47\times10^{-2}\;\mathrm{and}\;3.47\times10^{-2}, P\mathrm{s}<0.0001$, respectively). Similarly, sampling the \textit{multifractal linear} and \textit{IAAFT} noises from the center also canceled the otherwise observed doubling and quadrupling, respectively, of multiplicative cascades' narrowing of the right-side width with generations ($B\mathrm{s}=3.42\times10^{-2}\;\mathrm{and}\;4.56\times10^{-2}, SE\mathrm{s}=8.52\times10^{-3}\;\mathrm{and}\;8.52\times10^{-3}, P\mathrm{s}\;<0.0001$, respectively).

Incorporating center-sampled \textit{shuffled} noise canceled out the changes on the right-side width of the multifractal spectrum due to originally nonlinear multifractal noise, both on the $9$th generation and with progressively more generations ($B\mathrm{s}=-6.83\times10^{-2}\;\mathrm{and}\;1.89\times10^{-2}, SE\mathrm{s}=3.47\times10^{-2}\;\mathrm{and}\;8.52\times10^{-3}, P\mathrm{s}=0.049\;\mathrm{and}\;= 0.027$, respectively). An alternate regression only on the cascades with \textit{shuffled} noise revealed no such significant effects of center- versus left-sampling of the noise process at $p<0.05$. Hence, this finding is largely artifactual of a simultaneous contrast among cascades, the rest of which had expectable trends---both the linear trends induced by incorporating \textit{multifractal linear} and \textit{IAAFT} noises and the nonlinear trends implicit in originally nonlinear multifractal noise. So, these effects of center-sampled \textit{shuffled} noise indicate that multiplicative cascades with left-sampled \textit{multifractal linear} and \textit{IAAFT} noises make a stronger contrast with cascades incorporating left-sampled originally nonlinearly multifractal noise, and the contrast between originally nonlinear multifractal noise effects and shuffled noise effects on right-side width is just so small that it only becomes visible when we center-sample the \textit{multifractal-linear} and \textit{IAAFT} noises and thus diminish the effect of beginning-transient trends. Hence, multiplicative interactions across generations predicted initially greater right-side width and its subsequent narrowing with generations, and transient trends in progressively more linear, less multifractal noise (i.e., with progressive removal of the phase structure and then removal of the original spectrum width) accentuated both the initial widening and subsequent narrowing with generations. In brief, the linearization of \textit{multifractal linear} and \textit{IAAFT} noises induces a trend at the beginning of the noise process that diminishes the right-side width. If this effect complements the persisting or growing left-side width, these effects on the right-side width of the multifractal spectrum could indicate that linear trends across generations of a cascade process could make the spectrum more asymmetric.

The multiplicativity and type of noise at each generation appeared to influence the modal singularity strength, as indicated by the horizontal location of the peak of the multifractal spectrum, $\alpha(q=0)$. This peak was slightly less than $1$ for additive cascades with \textit{multifractal nonlinear} noise ($B=9.05\times10^{-1}, SE=1.54\times10^{-2}, P<0.0001$) and more so with left-sampled \textit{IAAFT} noise ($B=9.00\times10^{-2}, SE=2.17\times10^{-2}, P<0.0001$). It increased as well with multiplicative cascades applying left-sampled \textit{multifractal nonlinear} noise ($B=4.97\times10^{-1}, SE=2.17\times10^{-2}, P<0.0001$) and---to a greater degree---left-sampled \textit{shuffled} noise ($B=2.07\times10^{-1}, SE=3.07\times10^{-2}, P<0.0001$), but reduced with multiplicative cascades applying left-sampled \textit{multifractal linear} noise ($B=-1.57\times10^{-1}, SE=3.07\times10^{-2}, P<0.0001$) and left-sampled \textit{IAAFT} noise ($B=-5.84\times10^{-1}, SE=3.07\times10^{-2}, P<0.0001$). Hence, as with the right-side width of the multifractal spectrum, preserving the multifractal spectrum width in left-sampled \textit{multifractal linear} noise helps mitigate the decay of modal singularity strength attributable to the randomization of phase. This location of the peak increased with successive generations of multiplicative cascades with left-sampled \textit{multifractal nonlinear} noise ($B=3.10\times10^{-2}, SE=6.03\times10^{-3}, P<0.0001$), increasing with generations almost twice as fast with left-sampled \textit{multifractal linear} noise ($B=2.22\times10^{-2}, SE=8.52\times10^{-3}, P=0.009$) and more than twice as fast with left-sampled \textit{IAAFT} noise ($B=8.37\times10^{-2}, SE=8.52\times10^{-3}, P<0.0001$). Hence, multiplicative interactions across successive generations counteract the earlier effects of phase randomization, effectively building back the modal singularity strength of the cascade.

Again, center-sampling \textit{IAAFT} noise and \textit{multifractal linear} noise canceled out the effects noted above from left-sampled effects. Center-sampling \textit{IAAFT} noise canceled out the increase in the singularity of the peak in additive cascades ($B=-9.97\times10^{-2}, SE=3.07\times10^{-2}, P=0.001$), canceled out the reduction in peak singularity for multiplicative cascades ($B=6.70\times10^{-1}, SE=4.35\times10^{-2}, P<0.0001$), and canceled out the increase in peak singularity ($B=-7.69\times10^{-2}, SE=1.21\times10^{-2}, P<0.0001$). Similarly, center-sampling \textit{multifractal linear} canceled out the reduction in peak singularity for multiplicative cascades ($B=1.82\times10^{-1}, SE=4.35\times10^{-2}, P<0.0001$), and canceled out the increase in peak singularity ($B=-3.62\times10^{-2}, SE=1.21\times10^{-2}, P=0.003$).

With the foregoing narrowing of the right-side width, the peak $\alpha(q=0)$ value increase goes hand in hand with an extension of the left-side width of the multifractal spectrum. An important point to highlight is that, whereas the peak location increased one of the additive cascade types, the left-side width changed only for multiplicative cascades. Much like the horizontal location of the multifractal spectrum peak, $\alpha(q=0)$, the left-side width at $9$th generation increased for the multiplicative cascades with \textit{nonlinear} noise ($B=8.49\times10^{-1}, SE=2.17\times10^{-2}, P<0.0001$) and even more with \textit{shuffled} noise ($B=2.51\times10^{-1}, SE=3.07\times10^{-2}, P<0.0001$). However, the left-side width increased less for multiplicative cascades with \textit{multifractal linear} noise ($B=-1.59\times10^{-1}, SE=3.07\times 10^{-2}, P<0.0001$) and progressively less for multiplicative cascades with \textit{IAAFT} noise ($B=-5.82\times10^{-1}, SE=3.07\times 10^{-2}, P<0.0001$). Here, we see a major difference between the results for left-side width and those for the prior two features (i.e., right-side width and peak). In the prior features, \textit{multifractal linear} and \textit{IAAFT} noise had accentuated the effects of \textit{multifractal nonlinear} noise in multiplicative cascades. However, here on the left-side width, \textit{multifractal linear} and \textit{IAAFT} noise acts in the opposite direction from \textit{multifractal nonlinear} noise. So, phase-randomization reduces the left-side width. Yet again, even the opposition of these noises to the \textit{nonlinear} case shows a similar buffering of cascades' multifractal spectrum width by noise spectrum width. Although phase-randomized noise counteracts more of the nonlinear noise's effect on the left-side width, phase-randomized noise maintaining the multifractal spectrum width from the originally nonlinear case exerts less of this diminution of multifractal structure. Beyond the $9$th generation, the left-side width increased with successive generations for \textit{multifractal nonlinear} noise ($B=5.04\times10^{-2}, SE=6.03\times 10^{-3}, P<0.0001$). This increase was faster---by almost an additional half of the \textit{multifractal nonlinear} rate---for the multiplicative cascades with \textit{multifractal linear} noise ($B=2.20\times10^{-2}, SE=8.52\times 10^{-3}, P=0.010$), and the left-side width increased at almost double the \textit{multifractal nonlinear} rate ($B=8.66\times10^{-2}, SE=8.52\times 10^{-3}, P<0.0001$). So, again, as with the other two features of the multifractal spectrum, the successive multiplicative interactions across generations balance out the effects of the phase-randomized noises, both \textit{multifractal linear} and \textit{IAAFT}.

Once more, center-sampling \textit{IAAFT} noise and \textit{multifractal linear} noise canceled out the effects noted above from left-sampled effects. Center-sampling \textit{IAAFT} noise canceled out the reduction in the left-side width for multiplicative cascades ($B=1.79\times10^{-1}, SE=4.35\times10^{-2}, P<0.0001$), and canceled out the increase in peak singularity ($B=-3.69\times10^{-2}, SE=1.21\times10^{-2}, P=0.002$). Similarly, center-sampling \textit{multifractal linear} canceled out the reduction in peak singularity for multiplicative cascades ($B=7.03\times10^{-1}, SE=4.35\times10^{-2}, P<0.0001$), and canceled out the increase in peak singularity ($B=-8.22\times10^{-2}, SE=1.21\times10^{-2}, P<0.0001$).

\subsection{Hypothesis 2: Multiplicativity amplifies multifractal nonlinearity}

Multifractal spectrum width for the original series ($\Delta\alpha_\mathrm{Orig}$) was largely constant across generations $9$th through $15$th (Figs.~\ref{fig: f2} \& \ref{fig: f3}, \textit{top}), while multifractal spectrum widths for the linearly structured IAAFT surrogates ($\Delta\alpha_\mathrm{Surr}$) decreased (Figs.~\ref{fig: f2} \& \ref{fig: f3}, \textit{middle}). Specifically, $t$-statistic expressing multifractal nonlinearity by comparing $\Delta\alpha_\mathrm{Orig}$ and $\Delta\alpha_\mathrm{Surr}$---that is, $t_\mathrm{MF}$, grew across generations, reaching $t_\mathrm{MF}=50$ by the $15$th generation (Fig.~\ref{fig: f3}, \textit{bottom}). Hence, evidence for the cascade process's nonlinearity becomes clearer as the surrogate-spectrum width narrows with generations.

To model the change of these quantities, we used a full-factorial regression model, much like that for features above, replacing the covariate \textit{Feature} with an analogous covariate \textit{Outcome} to have baseline value of $\Delta\alpha_\mathrm{Orig}$ and alternate values $\Delta\alpha_\mathrm{Surr}$ and $t_\mathrm{MF}$. The highest-order interaction in the regression modeling of these quantities was Outcome $\times$ Type $\times$ Multiplicative $\times$ Generation, and modeling included all component lower-order interactions and main effects. In what follows, we detail the effects of these factors on $t_\mathrm{MF}$ both in terms of its continuous variation and its more dichotomous status as statistically significant or not, using the functions \texttt{lmer} and \texttt{glmer} from the package \texttt{lme4} \cite{bates2009package} in $R$ \cite{team2013r}.

\subsubsection{Multiplicative interactions across scales increased the continuous value of $t_\mathrm{MF}$ except when involving IAAFT noise}

The linear regression modeling of continuous $t_\mathrm{MF}$ had seven significant effects, all referring only to changes in $t_\mathrm{MF}$. The general indication was twofold. First, multiplicativity incorporating originally nonlinear multifractal noise promoted the growth of $t_\mathrm{MF}$ across progressive generations of the cascade. Second, whereas IAAWT-series and shuffled-series noises produced no difference in $t_\mathrm{MF}$ for multiplicative cascades as compared to originally nonlinear multifractal noise, multiplicative cascades with IAAFT noise reduced progressively diminished $t_\mathrm{MF}$ across progressive generations of the cascade. The sampling issue only appeared in the first $9$ generations, with left-sampling showing a diminution of $t_\mathrm{MF}$ in additive cascades and generating relatively larger $t_\mathrm{MF}$ in multiplicative cascades.

Results replicated various points from prior findings \cite{mangalam2024multifractalnonlinear}. For instance, by the $9$the generation, multiplicative cascades initially exhibited smaller $t_\mathrm{MF}$ ($B=-9.16\times10^{0}, SE=3.00\times10^{0}, P=0.002$). Subsequent generations of reverse this difference: additive cascades exhibited significantly lower $t_\mathrm{MF}$ with progressive generations ($B=-1.90\times10^{0}, SE=5.88\times10^{-1}, P= 0.001$), and multiplicative cascades exhibited significantly higher $t_\mathrm{MF}$ with progressive generations ($B=7.90\times10^{0}, SE=8.31\times10^{-1}, P<0.0001$). The notable exception to this initial finding was a set of cascades with IAAFT noise. Additive cascades with IAAFT noise had dramatically lower $t_\mathrm{MF}$ on the $9$th generation ($B=-1.29\times10^{1}, SE=3.00\times10^{0}, P<0.0001$), but multiplicative cascades with IAAFT noise had dramatically higher $t_\mathrm{MF}$ on the $9$th generation ($B= 1.62\times10^{1}, SE=24.24\times10^{0}, P=0.0001$). Then, multiplicative cascades with IAAFT showed none of the rest of the multiplicative cascades' growth of $t_\mathrm{MF}$ with progressive generations, instead predicting a diminution of $t_\mathrm{MF}$ with progressive generations ($B=-2.88\times10^{0}, SE=1.18\times10^{0}, P=0.014$).

The only difference in the effects for cascades with center-sampled noise was for the IAAFT-noise case. Center-sampling the IAAFT noise canceled out the observed decrease in $t_\mathrm{MF}$ for additive cascades ($B=1.18\times10^{1}, SE=4.24\times10^{0}, P=0.005$) and canceled out the relative increase of $t_\mathrm{MF}$ across the $9$th-generations for multiplicative cascades ($B=-1.29\times10^{1}, SE=4.24\times10^{0}, P=0.005$). The only feature that persisted when the IAAFT noise was sampled from the center was the already-reported diminution of $t_\mathrm{MF}$ for progressively more generations of multiplicative cascades. To summarize, the effect of the series-beginning trend of the IAAFT noise was to reduce and inflate $t_\mathrm{MF}$ in early generations of additive and multiplicative cascades. Whether the noise was left- or center-sampled, incorporating IAAFT noise led the multiplicative cascades to \textit{reduce} $t_\mathrm{MF}$ with successive generations.

\subsubsection{Multiplicative interactions across scales increased the likelihood of statistically significant $t_\mathrm{MF}$ except when involving IAAFT noise}

To model the number of significant positive values of $t_\mathrm{MF}>2.04$, we used a mixed-effect logistic regression with the function \texttt{glmer} from the package \texttt{lme4} \cite{bates2009package} in $R$ \cite{team2013r}. Models replicating the same family of higher-order and component lower-order interactions and main effects did not converge. Therefore, we trimmed the model to test only the sum of three main effects and one interaction, namely, Type, Multiplicative, Generation, and Type $\times$ Generation, and found significant effects.

The pattern of significance we found for explaining positive, significant $t_\mathrm{MF}$ reflected the same themes from the continuous-valued model of $t_\mathrm{MF}$. First, there were lower logarithmic odds for significant $t_\mathrm{MF}$ ($B=-6.36\times10^{-1}, SE=1.63\times10^{-1}, P<0.0001$) early on, that is, in the $9$th generation of the multiplicative cascades. Additive cascades showed a significant reduction in logarithmic odds of a significantly positive $t_\mathrm{MF}$ with each generation following the $9$th generation ($B=-3.04\times10^{-1}, SE=2.79\times10^{-2}, P<0.0001$). With each generation after the $9$th generation, there was a significant increase in the logarithmic odds of a significantly positive $t_\mathrm{MF}$ for multiplicative cascades ($B=9.77\times10^{-1}, SE=4.13\times10^{-2}, P<0.0001$). Lastly, the additive cascades with IAAFT noise showed the lowest logarithmic odds of a significantly positive $t_\mathrm{MF}$ ($B=-1.51\times10^{0}, SE=1.86\times10^{-1}, P<0.0001$). Adding multifractal outcomes for the center-sampled cascades with those for the left-sampled cascades led this regression model to fail to converge. This regression model with failed convergence did show that cascades with center-sampled IAAFT canceled out the left-sampled IAAFT decrease in logarithmic odds of significantly positive $t_\mathrm{MF}$ ($B=1.85\times10^{0}, SE=2.62\times10^{-1}, P<0.0001$). However, running the same regression model on multifractal outcomes for the center-sampled cascades alone revealed that center-sampling did reverse the results uniquely for IAAFT noise while leaving all other effects intact from the left-sampled results. Specifically, only modeling the multifractal outcomes for the center-sampled cascades suggested an increase in logarithmic odds of significantly positive $t_\mathrm{MF}$ for cascades with the center-sampled IAAFT noise ($B=3.77\times10^{-1}, SE=1.86\times10^{-1}, P=0.043$).

We also applied this regression modeling to significantly negative $t_\mathrm{MF}$ values. The same regression model converged after removing the interaction of Multiplicativity $\times$ Generation and considering only the main effects. Hence, we could not replicate the previous finding that multiplicative cascades were more likely to produce significant negative $t_\mathrm{MF}$ with fewer generations but less likely to produce significant negative $t_\mathrm{MF}$ with progressively more generations \cite{mangalam2024multifractal}. However, we found in the present case that additive cascades with IAAFT noise had the greater logarithmic odds of producing a significantly negative $t_\mathrm{MF}$ ($B=1.77\times10^{0}, SE=1.81\times10^{-1}, P<0.0001$), although progressively more generations of additive cascades reduced the logarithmic odds of a significantly negative $t_\mathrm{MF}$ ($B=-7.06\times10^{-2}, SE=1.78\times10^{-2}, P<0.0001$). Across all generations---early or late, multiplicative interactions reduced the logarithmic odds of significant negative $t_\mathrm{MF}$ ($B=-2.49\times10^{0}, SE=1.33\times10^{-1}, P<0.0001$). Notably, this effect of generations weakened but not---even at the $15$th generation---wiped out the logarithmic odds of significant negative $t_\mathrm{MF}$ for additive cascades with IAAFT noise. On the other hand, the effect of multiplicativity was sufficient to predict that multiplicative cascades with IAAFT noise would have significantly lower logarithmic odds of a significantly negative $t_\mathrm{MF}$ than the additive cascades with IAAFT noise.

\subsubsection{Disentangling length and generation effects}

\begin{figure*}
    \includegraphics[width=5.25in]{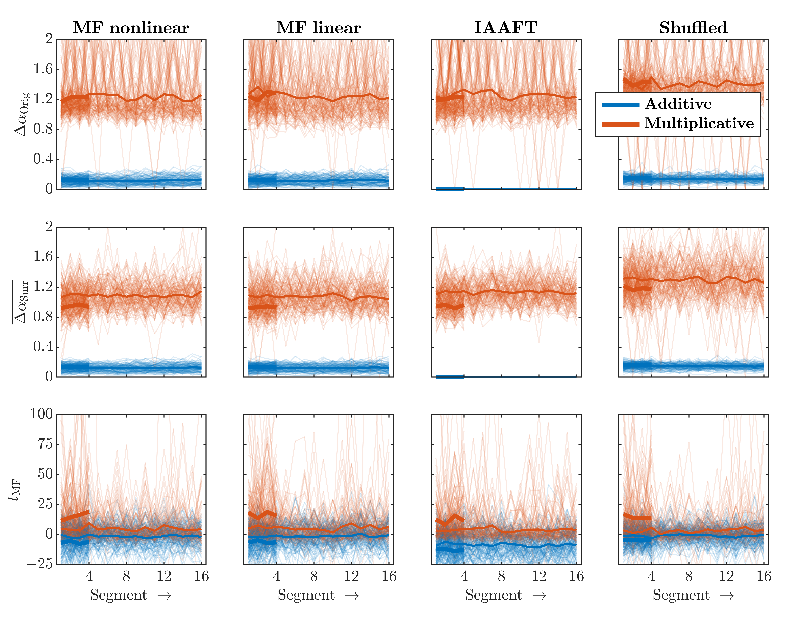}
    \caption{Multifractal spectral properties of the four additive and multiplicative cascade types across nonoverlapping segments in $15$th and final generation. Multifractal spectrum width for the original cascades $\Delta\alpha_\mathrm{Orig}$ (\textit{top}), multifractal spectrum width for the corresponding $32$ IAAFT surrogates $\overline{\Delta\alpha_\mathrm{Surr}}$ (\textit{middle}), and multifractal nonlinearity $t_\mathrm{MF}$ (\textit{bottom}). Thick and thin lines indicate segments of length $l=2^{12}\;\mathrm{and}\;2^{10}$, respectively. Each line represents the average values derived from $N=100$ numerical simulations.}
    \label{fig: f4}
\end{figure*}

\begin{figure*}
    \includegraphics[width=5.25in]{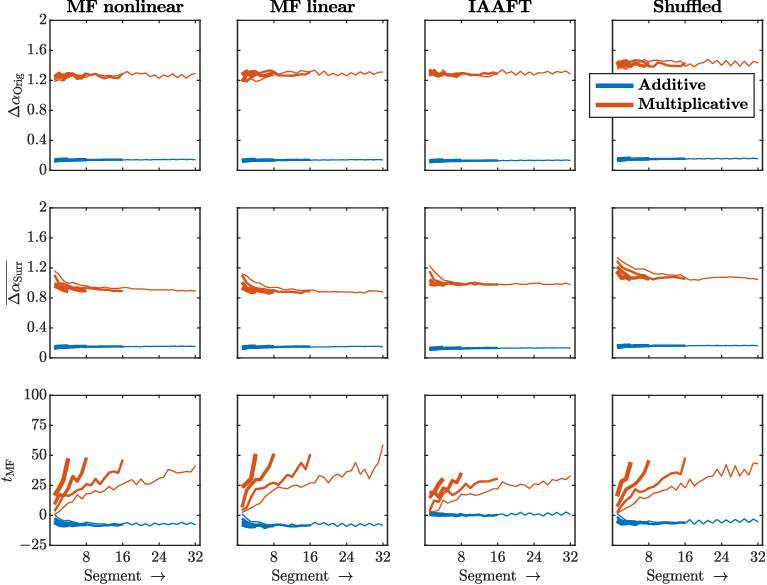}
    \caption{Multifractal spectral properties of the four additive and multiplicative cascade types across progressively longer segments from the beginning to the end in $15$th and final generation. Multifractal spectrum width for the original cascades $\Delta\alpha_\mathrm{Orig}$ (\textit{top}), multifractal spectrum width for the corresponding $32$ IAAFT surrogates $\overline{\Delta\alpha_\mathrm{Surr}}$ (\textit{middle}), and multifractal nonlinearity $t_\mathrm{MF}$ (\textit{bottom}). The line thickness decreases progressively from the longest segment length of $2^{12}$ to $2^{11}$, then to $2^{10}$, and finally to $2^{9}$. Each line represents the average values derived from $N=100$ numerical simulations.}
    \label{fig: f5}
\end{figure*}

\begin{figure*}
    \includegraphics[width=5.25in]{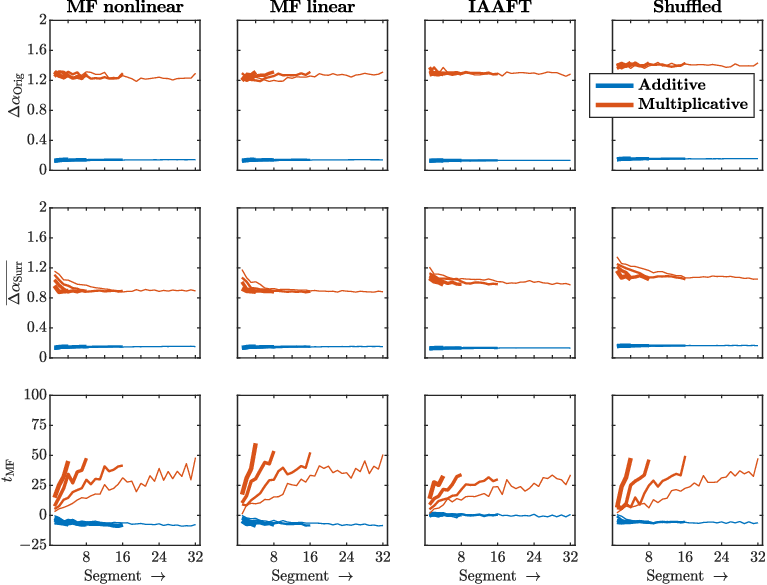}
    \caption{Multifractal spectral properties of the four additive and multiplicative cascade types across progressively longer segments from the end to the beginning in $15$th and final generation. Multifractal spectrum width for the original cascades $\Delta\alpha_\mathrm{Orig}$ (\textit{top}), multifractal spectrum width for the corresponding $32$ IAAFT surrogates $\overline{\Delta\alpha_\mathrm{Surr}}$ (\textit{middle}), and multifractal nonlinearity $t_\mathrm{MF}$ (\textit{bottom}). The line thickness decreases progressively from the longest segment length of $2^{12}$ to $2^{11}$, then to $2^{10}$, and finally to $2^{9}$. Each line represents the average values derived from $N=100$ numerical simulations.}
    \label{fig: f6}
\end{figure*}

The foregoing results provide encouraging evidence that multiplicative interactions across progressively more generations within a cascade strengthen the evidence of multifractal nonlinearity as indexed by both continuous $t_\mathrm{MF}$ and significantly positive $t_\mathrm{MF}$. However, it is worth noting that the canonical binomial-fracturing method of constructing cascades confounds the generation number with the length of the cascade time series. As noted in previous simulation work \cite{mangalam2024multifractal}, this confound can be demonstrated in two ways. First, we can take the final generations' resulting cascade time series and submit subsets of those final-generation time series to multifractal analysis and surrogate comparison. For instance, we can draw $2^{6},2^{5},2^{4},2^{3},2^{2},\;\mathrm{and}\;2^{1}$ nonoverlapping subsets of varying lengths, including $l=2^{8},2^{9},2^{10},2^{11},2^{12},\;\mathrm{and}\;2^{13}$, respectively, from the final-generation time series. However, the drawback of this first method is that the nonoverlapping definition of these subsets underestimates any effect of the abrupt shifts across binomial fracturing. So, we can examine progressively longer subsets whose lengths are not explicitly powers of $2$. Length dependencies across these nonoverlapping segments within the $15$th generation of the cascade model showed stable $\Delta\alpha_\mathrm{Orig}(q)$, decreasing $\Delta\alpha_\mathrm{Surr}(q)$, and growing $t_\mathrm{MF}$ (Fig.~\ref{fig: f4}) with longer nonoverlapping subset. Second, we can draw progressively longer subsets anchored on either the first or last values of the final-generation time series. More specifically, we can use natural-number multiples of our segment lengths (e.g., for $j+1$th generation of a cascade, $l=m\cdot 2^{7+s},m=1,2,3,\dots,n/2^{7+s}$, where $n$ is the time series length ($=2^{14}$) and $s\leq j-7$ and anchoring, to begin with, the first value of the series $p_{1,j+1}$ (e.g., spanning $[p_{1,j+1},p_{2,j+1},\dots,p_{m\cdot 2^{7+s},j+1}]$) or to end with the last value of the series (e.g., spanning $[x_{2^{j}-m\cdot2^{7+s}+1,j+1},x_{2^{j}-m\cdot2^{7+s}+2,j+1},\dots,x_{2^{j},j+1}]$). For these overlapping, natural-number-multiple-length subsets of the final time series, the relationships did not depend on whether subsets began from the beginning and grew towards the end of the $15$th generation of the cascade time series (Fig. ~\ref{fig: f5}) or from the beginning and grew towards the end of the $15$th and final generation of the cascade time series (Fig. ~\ref{fig: f6}). In both cases of overlapping subsets, $\Delta\alpha_\mathrm{Orig}(q)$ remained stable, $\Delta\alpha_\mathrm{Surr}(q)$ reduced, and $t_\mathrm{MF}$ grew with greater length.

\begin{figure*}
    \includegraphics[width=5.25in]{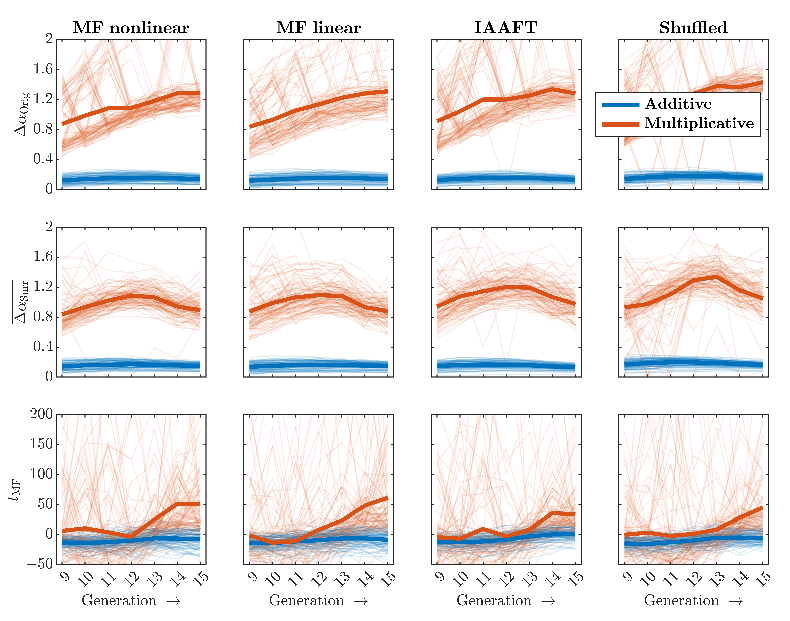}
    \caption{Multifractal spectral properties of the four additive and multiplicative cascade types across successive generations after controlling for length. Multifractal spectrum width for the original cascades $\Delta\alpha_\mathrm{Orig}$ (\textit{top}), multifractal spectrum width for the corresponding $32$ IAAFT surrogates $\overline{\Delta\alpha_\mathrm{Surr}}$ (\textit{middle}), and multifractal nonlinearity $t_\mathrm{MF}$ (\textit{bottom}). To ensure consistent length across generations, time series were padded with consecutive repetitions of individual-cell values for $9$th through $15$th generations $64,32,16,8,4,2,\;\mathrm{and}\;1$ times, respectively. The bold lines represent the average values derived from $N=100$ numerical simulations, while the finer lines alongside them portray individual data points within these simulation sets.}
    \label{fig: f7}
\end{figure*}

We had previously simulated ``padded'' versions of the cascade time series to break the length-generation confound. Effectively, this strategy ensures that all series are the same length, and it merely defines the cells of the binomial-fracturing cascade process over progressively fewer nonoverlapping subsets of the same series length. Hence, the length of the series is the same for every generation. In contrast, previously, a parent cell might have given rise to two children's cells in the next generation. This padding entailed that each parent was simply a subset of the series double the length of each child's cell. Another way to think about this procedure is in terms of the MATLAB procedure ``repelem'' that takes two inputs (a vector $a$ of length $n_{a}$ and scalar $n_{b}$ such that $n_{b}/n_{a}$ is a natural number) and outputs a new vector $b$ of length $b$ for which each $i$th element of $a$ becomes the $(i-1)*(n_{b}/n_{a})+1$st,$(i-1)*(n_{b}/n_{a})+2$nd,$\dots$,$(i-1)*(n_{b}/n_{a})+(n_{b}/n_{a})-1$th,$(i-1)*(n_{b}/n_{a})+(n_{b}/n_{a})$th values of $b$. Removing the ``padding'' of repeated values is necessary to compute a surrogate for such a series. The repeated values make it all too likely that a phase randomization that preserves the amplitude spectrum for the padded series would have a compared multifractal spectrum. However, multifractal analysis could proceed for the padded original and surrogate series. What this procedure accomplishes is to produce equally long series that reflect different numbers of accumulated interactions across time scales. When we thus control for length, the outcome is comparable once more to all foregoing examples: $\Delta\alpha_\mathrm{Orig}(q)$ remained stable, decreasing $\Delta\alpha_\mathrm{Surr}(q)$, and $t_\mathrm{MF}$ grew with greater numbers of generations (Fig.~\ref{fig: f7}). Hence, we confirm that the relations we found were not exclusively specific to length.

\subsection{Hypothesis 3: Multifractal diversity influences ergodicity breaking}

Cascades consistently broke ergodicity, a phenomenon detailed in prior research \cite{mangalam2024multifractal}. Interestingly, multiplicative and additive cascades exhibited distinctive forms of ergodicity breaking. Multiplicative cascades, known for generating non-Gaussian histograms (e.g., \cite{kiyono2007estimator, kiyono2008non}), yielded higher intercepts on the $EB$-vs.-$t$ curves compared to their additive counterparts. In stark contrast, additive cascades displayed notably shallow $EB$-vs.-$t$ curves, especially evident as the cascades progress from generations $9$ through $15$ (Fig.~\ref{fig: f7}, \textit{top}), emphasizing the robust presence of ergodicity breaking within the additive system. Intriguingly, examining the $EB$-vs.-$t$ curves of additive cascades revealed no distinction between those generated from multifractal nonlinear and multifractal linear components.

\begin{figure*}
    \includegraphics[width=5.25in]{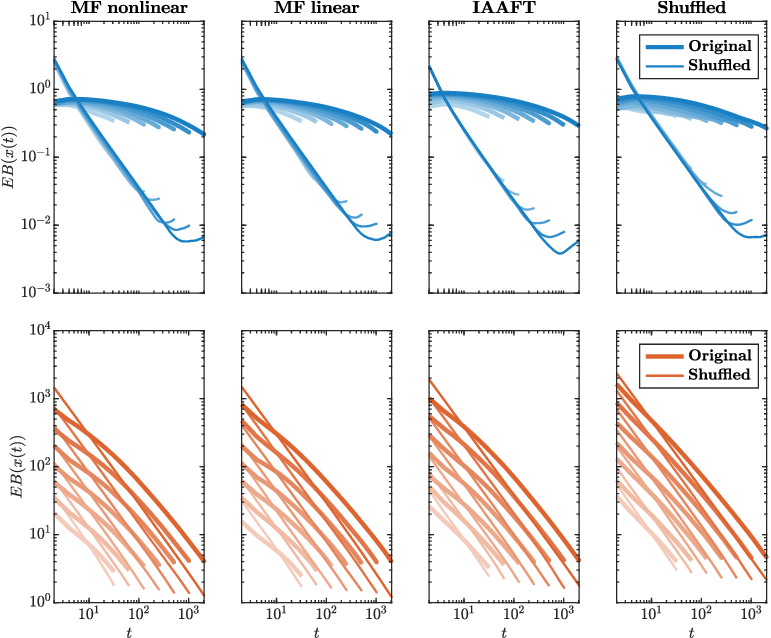}
    \caption{Ergodicity breaking in the four additive (\textit{top}) and multiplicative (\textit{bottom}) cascade types across successive generations. This phenomenon is quantified using the ergodicity breaking factor $EB(x(t))$ (lag $\Delta=10$ samples). The traces progressively deepen in color from $9$ through generation $15$. Notably, the time series length experiences exponential growth across these generations---$2^{8},2^{9},2^{10},2^{11},2^{12},2^{13},\;\mathrm{and}\;2^{14}$ for the $9$th through $15$th generation, respectively. Each line represents the average values derived from $N=100$ numerical simulations.}
    \label{fig: f8}
\end{figure*}

\begin{figure*}
    \includegraphics[width=5.25in]{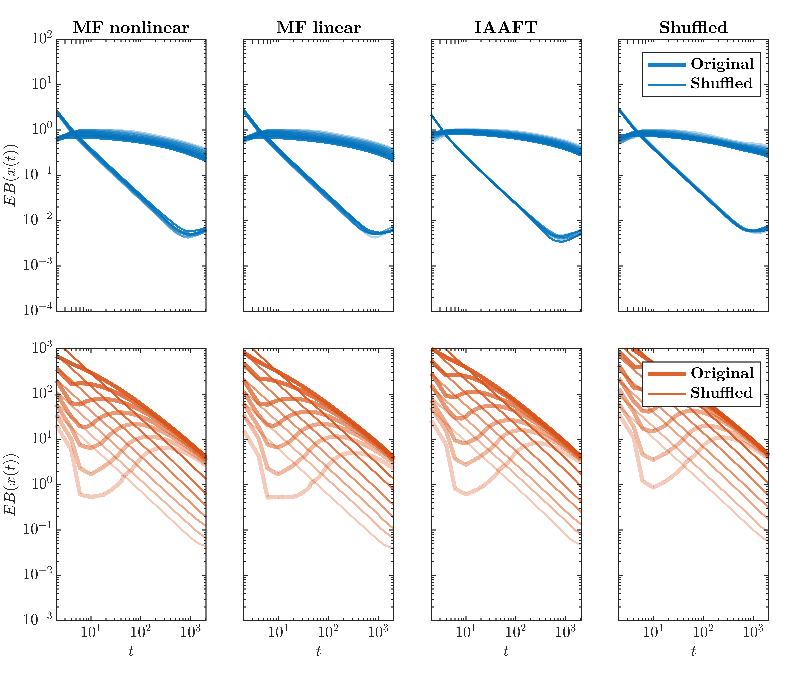}
    \caption{Ergodicity breaking in the four additive (\textit{top}) and multiplicative (\textit{bottom}) cascade types across successive generations after controlling for length. This phenomenon is quantified using the ergodicity breaking factor $EB(x(t))$ (lag $\Delta=10$ samples). The traces progressively deepen in color from $9$ through generation $15$. To ensure consistent length across generations, time series were padded with consecutive repetitions of individual-cell values for $9$th through $15$th generations $64,32,16,8,4,2,\;\mathrm{and}\;1$ times, respectively. Each line represents the average values derived from $N=100$ numerical simulations.}
    \label{fig: f9}
\end{figure*}

It is worth noting that the $EB$-vs.-$t$ curves assume flatter profiles for additive cascades constructed from phase-randomized IAAFT noise. Notably, neither the original nor the shuffled multiplicative cascades reach small and diminishing values for the ergodicity breaking factor ($EB$), as observed in shuffled additive cascades---however, the $EB$-vs.-$t$ curves are visibly shallower for the original time series compared to the shuffled time series, particularly in the cases of multiplicative cascades with originally nonlinear multifractal noise and linear multifractal noise of the same spectrum width. In contrast, the curves are steeper for multiplicative cascades with IAAFT and shuffled noise. This nuanced distinction is evident in the bottom row of Fig.~\ref{fig: f8}, where the original $EB$-vs.-$t$ curves (bolded-line traces) extend beyond two shuffled-series curves for multiplicative cascades with originally nonlinear and linear multifractal noise. For multiplicative cascades with IAAFT noise, the $EB$-vs.-$t$ curves barely cross two shuffled-series curves. Conversely, multiplicative cascades with shuffled noise exhibit a more linear decrease, crossing only one shuffled-series curve. This pattern holds even for padded ``repelem''-type cascades that control for length (Fig.~\ref{fig: f9}).

In summary, the $EB$-vs.-$t$ curves decay more slowly over shorter time scales for cascades with original nonlinear multifractal and IAAWT-specified linear multifractal noises. Although multiplicative cascades seem to display weaker ergodicity breaking than additive cascades, the comparable nature of ergodicity breaking across both cascade types suggests that the multifractal spectrum width crucially determines the extent of weak ergodicity breaking in the multiplicative case.

\section{Discussion}

This investigation unveils the outcomes of numerical simulations, underscoring the dominance of multiplicative interactions over the intrinsic nonlinear characteristics of constituent processes. To illuminate this phenomenon, we executed simulations involving cascade time series with component processes operating at disparate timescales. These processes were characterized by four distinct properties: multifractal nonlinearity, multifractal linearity from constrained phase-randomization (achieved through the Iterative Amplitude Adjusted Wavelet Transform applied to multifractal nonlinearity), phase-randomized linearity (achieved via the Iterative Amplitude Adjustment Fourier Transform), and phase- and amplitude-randomized (achieved through shuffling). Our results unequivocally establish that the emergent multifractal properties are decisively dictated by the multiplicative interactions among components rather than the intrinsic properties of the component processes themselves. Remarkably, even component processes exhibiting purely linear traits can engender nonlinear interactions across scales when these interactions assume a multiplicative nature. In stark contrast, additive interactions among component processes invariably yield linear outcomes. These findings provide a robust theoretical foundation for current interpretations of multifractal nonlinearity, firmly situating its origins in the realm of multiplicative interactions across scales within biological and psychological processes. We tested three primary hypotheses. First, we posited that multiplicative cascades fostering interactivity across timescales would generate wider multifractal spectra---both on the right and left sides, with peaks corresponding to greater singularity strength $\alpha$—compared to additive cascades. Second, we anticipated that successive generations of multiplicative cascades would yield larger and more statistically significant values of $t_\mathrm{MF}$, surpassing the relationships between length and number of generations inherent in cascade simulations. Third, we expected that multiplicative cascades would exhibit more pronounced ergodicity breaking based on non-Gaussianity and weaker sequence-driven ergodicity breaking than additive cascades. Throughout, building on our previous discovery that these effects of multiplicative cascade dynamics were more pronounced in cascades incorporating $fGn$ at each generation \cite{mangalam2024multifractal}, we explored whether these effects of multiplicativity depended on multifractal noise. Our results validated all these predictions, except that progressive generations of multiplicative cascades seemed to accentuate a decay of right-side spectrum width.

\subsection{Multifractal noise with nonlinear correlations and with linear transients each led multiplicative cascades to promote left-side asymmetry in multifractal spectra}

Incorporating multifractal noise processes into multiplicative cascades resulted in multifractal spectra characterized by a broader left side and a more singular peak (i.e., exhibiting a higher $\alpha(q = 0)$)---more asymmetric than previous findings involving monofractal $fGn$ in multiplicative cascades. The width of the multifractal spectrum on the right side aligned with prior research when multifractal noise was introduced, resembling outcomes observed with both additive and multiplicative cascades utilizing $awGn$ but not with $fGn$ \cite{mangalam2024multifractal}. Additive cascades progressively narrowed the right-side width of the multifractal spectrum with successive generations, while multiplicative cascades initially expanded this width at the 9th generation. These novel outcomes suggest that the application of multifractal \textit{nonlinear} noise or phase-randomized multifractal \textit{IAAFT} noise in early generations can either enhance or diminish, respectively, the heterogeneity represented by small fluctuations on the right side of the multifractal spectrum. Departing from earlier work, multifractal noises generally reduced the right-side width of the multifractal spectrum, with two nuances: firstly, multiplicative cascades with multifractal \textit{nonlinear} noise narrowed the right-side width in subsequent generations, and secondly, phase-randomized multifractal noises increased the $9$th generation's right-side width and accentuated narrowing, particularly with \textit{IAAFT} noise compared to the \textit{nonlinear} case. Confirming previous findings that linearity undermines small-fluctuation heterogeneity, our results indicate that preserving the original multifractal spectrum width will likely maintain small-fluctuation heterogeneity. Furthermore, multiplicative cascades, especially with monofractal $fGn$, may be more adept at sustaining greater right-side spectrum width across generations. The observed small-size heterogeneity may be attributed to the ergodicity-breaking capacity of $fGn$ and the ability of multiplicative interactions across time scales to translate these effects into heterogeneous sequences within smaller-sized fluctuations. Likewise, preserving the multifractal spectrum width for noise applied at each generation reinforces cascades with more robust multifractal spectrum widths, a factor not contributing to the multifractal spectrum widths of additive cascades.

This study also reveals how asymmetric multifractal spectra can indicate sensitivity to progressively linearized multifractal noise trends. Despite the initial generations featuring only a handful of children (e.g., $2,4,8,\;\mathrm{and}\;16$ in the first $4$ generations), the expanding pool of offspring in later generations amplifies the influence of diverse multifractal noises on the resulting multifractal structure. The robust time-symmetry of IAAFT noise exerts a profound impact, narrowing both sides of the multifractal spectrum and significantly reducing $t_\mathrm{MF}$ in additive cascades while concurrently boosting $t_\mathrm{MF}$ in the earliest generations of multiplicative cascades. Transient trends at the onset of IAAWT noise also lead to a similar constriction in the resulting cascades' multifractal spectra, albeit without any noticeable effects on $t_\mathrm{MF}$. This observation is intriguing, considering that IAAWT noise represents a linearization that alters the original sequence while retaining the same multifractal spectrum—a constrained form of phase-randomization \cite{keylock2018gradual}. Even this constrained phase randomization proves sufficient to generate transients that multiplicative cascades manifest through substantial alterations in the multifractal spectrum. Shuffled multifractal noise exhibits no statistically discernible effects until all these noise processes are sampled from the center. In this scenario, cascades involving the two linearizations of multifractal noise (IAAFT and IAAWT) exhibit the weakest trends, displaying minimal deviation from the originally nonlinear multifractal noise. Within these conditions, multiplicative cascades with shuffled multifractal noise showcase their ability to maintain right-side width. In contrast, multiplicative cascades incorporating originally nonlinear multifractal noise yield more asymmetric multifractal spectra with broader left sides. Meanwhile, even when the center of IAAFT processes is sampled, multiplicative cascades with IAAFT noise lead to a progressive reduction of $t_\mathrm{MF}$ in subsequent generations.

To summarize the effects on spectral features, multifractal noise led to progressively more asymmetric multifractal spectra with greater left-size width, and this asymmetry grew with successive generations for originally nonlinear multifractal noise and left-sampled linear surrogates of multifractal noise. Hence, incorporating multifractal noise can skew the multifractal spectrum of the resulting cascade even with nonlinear correlations in the noise, and linear trends in the noise process unfolding across each generation might accentuate that skewness. \textit{So, when behavioral and biological sciences find asymmetric multifractal spectra for any single observable, important questions to explore involve (i) whether the observable is receiving multifractal noise from a component process, (ii) whether there are nonlinear and linear sources of such noise, and (iii) whether the linear components of this multifractal noise are in full swing or just beginning to participate in cascade-like relationships across the organism's degrees of freedom.} It would also be important to consider how initiating a linear constraint on multifractal noise on the organism's other degrees of freedom might induce a transient asymmetry in the resulting multifractal spectra.

\subsection{Multiplicative cascades promoted greater multifractal nonlinearity $t_\mathrm{MF}$ except with IAAFT noise}

We observed intriguing patterns in the emergence of multifractal nonlinearity across generations. Notably, generations characterized by a greater number of scales exhibited a higher value of $t_\mathrm{MF}$ when comparing the original multifractal spectrum with the spectrum of linear surrogates. This finding aligns with prior empirical research that interprets increased multifractal nonlinearity as indicative of more intricate interaction-driven dynamics supporting perception and action (as reviewed in \cite{kelty2022turing}). As the data's sampling duration increased, the growth in multifractal nonlinearity became apparent, providing more robust evidence of nonlinear interactions across scales via IAAFT surrogate comparison, suggesting that nonlinear evidence of interactions intensifies with the time series length. Importantly, the independence of multifractality from specific sampling times enhances its reliability as an empirical technique. Multifractal nonlinearity remains constant within a data segment sampled for a specific duration, implying that any chunk of data can represent the entire process, provided the sampling length remains consistent across repeated sampling or comparisons across groups or conditions. It is crucial to emphasize that shifting values from the start to the end does not impact when the process is sampled but depends solely on the extent of data sampled. This consistency is vital for interpreting results from extended datasets, ensuring alignment with analyses of datasets excluding later data. Understanding that subsequent events cannot retroactively change the reality of prior evolution in a physical system underscores the significance of consistent interpretation, particularly when analyzing extended datasets. The observed consistently multifractal nonlinearity across same-length chunks along the final cascades reinforces this point, particularly in the context of multiplicative interactions, supporting our central finding on the pivotal role of such interactions in shaping multifractal emergent properties. Center-sampling only nullified IAAFT's effects on additive cascade $t_\mathrm{MF}$ and early-generation $t_\mathrm{MF}$ in multiplicative cascades. Regardless of controlling for transient trends at the beginning of IAAFT noise, multifractal cascades consistently exhibited smaller and less often significant values of $t_\mathrm{MF}$. Hence, the asymmetry of multifractal spectra due to all but shuffled-noise multiplicative cascades does not alter the outcome of most multiplicative cascades' multifractal nonlinearity. Comparing the entire cascade series to its own IAAFT surrogates allowed controlling for many linear factors in left-sampled noises. Whether sampled from the left or the center of the noise process, IAAFT noise consistently prevented multiplicative cascades from promoting multifractal nonlinearity across generations.

\subsection{Noise with greater multifractal spectrum with contributed to stronger sequence-dependent ergodicity breaking in multiplicative cascades}

Finally, our findings regarding ergodicity breaking are particularly thought-provoking. They add confirmatory evidence supporting both the previous theorizing that ergodicity breaking might stem from the sequential structure as well as from failures of processes to stabilize around Gaussian mean \cite{lebowitz1973modern} \textit{and} the more recent evidence that additive and multiplicative cascades distinctly promote the former and latter sources, respectively \cite{mangalam2024multifractal}. We found that additivity promoted sequence-driven ergodicity breaking, as in the case of the linear process $fGn$ showing strong ergodicity breaking with greater temporal correlations \cite{deng2009ergodic}. We also found that multiplicative cascades showed less sequence-driven ergodicity breaking but greater ergodicity breaking due to non-Gaussian histogram. Again, as in \cite{mangalam2024multifractal}, noise type affected sequence-driven ergodicity-breaking that depended on multiplicativity: the entailment of more homogeneous, more time-symmetric temporal structure in the IAAFT-noise promoted sequence-driven ergodicity breaking in additive cascades, but the wide multifractal spectrum in nonlinear multifractal noise and its preservation in the linear multifractal noise served to strengthen sequence-driven ergodicity breaking in multiplicative cascades. 

Hence, the assessment of ergodic properties remains a pertinent endeavor when modeling biological and psychological processes \cite{kelty2022fractal, kelty2023multifractaldescriptors, mangalam2021point, mangalam2022ergodic, mangalam2023ergodic, mangalam2023temporal}. And it now appears consistently that both the mathematical relationship across generations (i.e., multiplications or additions) and the temporal structure in the noise implicated in the cascade process (e.g., originally nonlinear multifractal, IAAWT, IAAFT, or shuffled) could each play a part in determining the ergodicity-breaking properties of our raw measurements. Beyond replicating this point from our previous simulation work \cite{mangalam2024multifractal}, an intriguing implication of the present findings is that IAAFT noise can provoke strong sequence-driven ergodicity breaking in additive cascades but diminish it in the multiplicative cascades. A complimentary intriguing implication is that a wider multifractal spectrum appears to promote sequence-driven ergodicity breaking, and the comparable ergodicity-breaking results for multiplicative cascades with originally nonlinear multifractal noise and IAAWT noise suggests that nonlinear correlations might not themselves be necessary for sequence-dependent ergodicity breaking. Then again, the contrast of results between cascades with originally nonlinear multifractal noise and cascades with IAAFT noise suggests that nonlinear correlations certainly suffice to produce a difference in ergodicity-breaking. Although the failure of mixing (i.e., the linear correlations in IAAFT noise) is sufficient for sequence-driven ergodicity breaking in additive cascades, the greater multifractal spectra for cascades with originally nonlinear multifractal noise entail \textit{variety} in the failure of mixing. Previous work had shown that multiplicative cascades with non-mixing $fGn$ indeed showed this greater sequence-dependent ergodicity breaking \cite{mangalam2024multifractal}; now we learn that multiplicative cascades can show greater sequence-dependent ergodicity with greater varieties in failure of mixing. In summary, we conclude that multiplicative cascades might protect against sequence-dependent ergodicity breaking, bringing relative stability to the cascade outcome precisely when the underlying noise process shows the least variety in its failure of mixing. The same multiplicative cascades appear to unleash relatively more sequence-dependent ergodicity breaking when the underlying noise processes are more variable in their failure of mixing.

\subsection{Conclusion and future directions}

We showed in this simulation work that multiplicative formalisms represent the most robust approach for evaluating whether fluctuations in empirical time series indicate multiplicative interactions among component processes operating across various scales and in various parts of an extended system. Previous work investigating cascade-driven simulations has traditionally only addressed isolated cascade processes, essentially isolating the parameters of a single cascade and confirming which range of parameter values could produce specific ranges of fractal and multifractal outcome measures. This work has been crucial for all empirical work considering any observable that may embody a single cascade process. As noted in the Introduction, simulation work with single cascades provides an ample testing ground for exploring why and under what constraints we should expect multifractal outcomes with or without sequence-driven ergodicity breaking---but only in single or isolated degrees of freedom. So long as we hope for cascade-dynamical modeling to speak to organism-wide interactivities, the next crucial step in cascade modeling may be to derive valid theoretical predictions for cascades that are not simply producing multifractal noise but incorporating different forms of multifractal noise as well. The present results provide new perspectives for interpreting how asymmetric multifractal spectra in single observables might reflect multiplicative cascades operating upon multifractal noise absorbed elsewhere in the organism or task context. This theoretical work allows asking more subtle questions about the proposed linearity or nonlinearity implicit in multifractal fluctuations endogenous to the organism (e.g., \cite{mangalam2020global, mangalam2020multifractal}) or in exogenous multifractal stimulation from the task context (e.g., \cite{kelty2023multifractalstimulation, stephen2011strong, ward2018bringing}).

This proposed direction for theoretical work is decidedly not a program of building multifractal noise to provide a ``turtles all the way down'' explanation of the origin of multifractal structure in biological and psychological measurements. First, we do not expect that cascade dynamics should be purely transparent to multifractal perturbation---that is, cascade dynamics themselves may be a fundamental mechanism sufficient to produce multifractal structure, and cascade dynamics is a compelling mechanism because cascade-like fracturing is not in and of itself a multifractal mechanism. In other words, cascades constitute minimally only one straightforward route for producing multifractal structure, and there is nothing inherently multifractal about the repetition of biased splitting of proportions in parent cells \cite{halsey1986fractal}. Second, we do not aim to re-present long-understood rules of superposition by which adding one fractal signal to another might produce a new signal of superposed fractality \cite{hu2001effect}, and the present work is not simply adding one multifractal time series to another multifractal time series of comparable length. Beyond what prior research has shown about potentially multifractal or nonmultifractal routes to multifractal results, we can point to the present results as evidence of how diverse the effects of incorporating multifractal noise can be. Indeed, the capacity for zero and (significantly) negative $t_\mathrm{MF}$ in the additive cascades and even of multiplicative cascades with IAAFT noise suggests that cascades chart many routes towards nonmultifractal results. While it may not seem significant at first glance, IAAFT noise consistently demonstrated a notable reduction in multifractal spectrum width compared to the original nonlinear multifractal noise. However, the persistence of a residual non-zero multifractal spectrum width in the histogram skew technically qualifies IAAFT noise as multifractal. Furthermore, our observations reveal a noteworthy phenomenon: numerous additive cascades featuring multifractal noise may present only scant evidence of multifractal structure, as exemplified by the subtle traces in the \textit{blue lines} across all \textit{top-row panels} in Figs.~\ref{fig: f3}, \ref{fig: f4}, \ref{fig: f5}, and \ref{fig: f6}. Consequently, no guarantee cascades will consistently yield compelling evidence of multifractality. Similarly, there is no categorical assurance that we should exclusively expect multifractal structure from cascades generated by applying multiple generations of multifractal noise.

Rather, cascades that incorporate multifractal noise open up a field of theoretical inquiry that paves a path towards scaling a multifractal model of biological and psychological processes up from single observables to an entire organism---that path that network science has already begun to hint at \cite{xiao2021deciphering}. The present exploration of cascades with multifractal noise is an early step in learning how to interpret our single observables' multifractality better. The present work does not give indicators of network topology; instead, it offers a view of how classic categories of cascades (i.e., additive and multiplicative) react to and involve multifractal noise that might spread through the system from spatiotemporally neighboring points of cascade dynamics. This work informs this line of theoretical inquiry by indicating that the mathematical form of cascade dynamics (i.e., additive or multiplicative) radically changes the effect of the simulated observable absorbing multifractal noise. For instance, we confirmed a previous finding that multiplicative cascades show stronger $t_\mathrm{MF}$ signatures of nonlinearity with a greater number of generations \cite{mangalam2024multifractal}. However, multifractal cascades generate much stronger $t_\mathrm{MF}$ signatures of nonlinearity with multifractal noise than with less multifractal IAAFT noise. This point extends the previous finding that multiplicative cascades involving $fGn$ noise showed greater $t_\mathrm{MF}$ than multiplicative cascades with $awGn$ \cite{mangalam2024multifractal}. Hence, we may be able to rank order the strength of multifractal nonlinearity from greatest to least for multiplicative cascades involving multifractal noise with wider spectra or more nonlinear correlations, multifractal noise with narrower spectra more closely approximating monofractal $fGn$ as well as $awGn$. This work also informs this approach by indicating that noise with greater multifractal spectrum width (e.g., in the comparison between originally nonlinear multifractal noise and IAAWT noise) and greater nonlinear correlations (i.e., in the contrast between originally nonlinear multifractal noise and IAAFT noise) can promote specifically different forms of ergodicity breaking.

Hence, the present simulation is a critical step in establishing an interactive framework to understand the adaptive functions of organisms, for instance, perceiving, acting, and cognizing. Instead of presenting a novel concept, we have brought Turing's age-old wisdom to modeling how we might interpret multifractal results when expecting our cascade-like measurements to bear the multifractal imprints of cascade processes elsewhere within the same body. The next steps for this process involve elaborating beyond the strictly additive and strictly multiplicative (e.g., \cite{mangalam2024multifractalnonlinear}) and also developing the structure of multifractal noise to simulate different network relationships feeding into the cascade simulation (see also \cite{kelty2023multifractalstimulation}). We hope also to explore the possibility that multifractally nonlinear signatures of cascade dynamics might help to classify qualitatively different modes of biological motility \cite{mangalam2023ergodic}. Behind this prospect is the further goal to connect our work with the network modeling that already provides glimmers of the relationship between distributed coordination and multifractal cascades \cite{xiao2021deciphering}. A longer-range goal of this work would be to develop and empirically test predictions of how well multifractal nonlinearity can detect changes in cascade structure due to endogenous multifractal behavior of the organismal network, especially as it may change with network topology or with successful goal satisfaction. At that point, it might even be possible to begin sketching out methods for simulating exogenous perturbations that could begin to develop theoretical predictions for how external threats or pressures might reshape the cascades that organismal networks of observables wield. Such theoretical advances would be welcome support for the growing empirical attempt to elaborate so-called ``stochastic resonance'' or ``noise-based'' stimulation beyond the classically ``white-noise'' $awGn$ structure \cite{collins1996anoise, collins1996bnoise, collins1997noise, galica2009subsensory, mcdonnell2009stochastic, mcdonnell2011benefits, miranda2016sensory, priplata2002noise, priplata2003vibrating} to more endogenously naturalistic fractal and multifractal types of stimulation \cite{hove2012interactive, kelty2023multifractalstimulation, nozaki1999mechanism}. To summarize, the simulations herein furnish a sturdy theoretical foundation that aligns with current interpretations of multifractal nonlinearity and can root further work to articulate the cascade dynamics spanning diverse scales within biological and psychological processes.\\

\noindent{}\textbf{Acknowledgments:} This work was supported by the Center for Research in Human Movement Variability at the University of Nebraska at Omaha, funded by the NIH award P20GM109090.\\

\noindent{}\textbf{Competing interests:} The authors have no conflict of interest.\\

\bibliography{apssamp}

\end{document}